\documentclass[final,5p,times,twocolumn,number,sort&compress]{elsarticle}

\usepackage{amssymb}
\usepackage{lipsum}

\usepackage{amsmath}
\usepackage{adjustbox}
\usepackage{graphicx}

\usepackage{booktabs}       

\usepackage{amsfonts}       
\usepackage{nicefrac}       
\usepackage{microtype}
\usepackage{hyperref}
\usepackage{cleveref}
\usepackage{xcolor}
\usepackage{multirow}
\usepackage{makecell}
\usepackage{array}
\usepackage{orcidlink}
\usepackage{float}
\usepackage{siunitx}
\usepackage{tabularx}
\usepackage{array}

\newcommand{\red}[1]{\textcolor{red}{#1}}

\journal{Physics Letters B}
\begin{document}

\begin{frontmatter}
\title{Amortized Simulation-Based Inference of Relativistic Mean-Field Couplings for Neutron-Star Equations of State}
\author[yonsei]{Prashant Thakur}
\author[coimbra]{Tuhin Malik}
\affiliation[yonsei]{
    organization={Department of Physics, Yonsei University},
    city={Seoul},
    postcode={03722},
    country={Republic of Korea}
}
\affiliation[coimbra]{
    organization={CFisUC, Department of Physics, University of Coimbra},
    city={Coimbra},
    postcode={3004-516},
    country={Portugal}
}
            
\begin{abstract}
We present a simulation-based inference framework for constraining microscopic relativistic mean-field parameters of neutron-star equations of state. Neural posterior estimation is applied to two representative RMF families, a density-dependent DDB model and a nonlinear RMF-NL model, using nuclear saturation properties, chiral effective-field-theory pure-neutron-matter pressures, and the maximum-mass constraint as conditioning observables. The inferred posteriors are validated against the conventional nested sampler (PyMultiNest) calculations and tested with the TARP coverage diagnostic. For both RMF parametrizations, the neural posterior reproduces the nested-sampling constraints on model couplings, nuclear-matter properties, and neutron-star observables with no significant bias. The amortized estimator generates $3\times 10^{4}$ posterior samples in about $2.5\,\mathrm{s}$ on a CPU, enabling a rapid inference workflow without the need for retraining for updated data. This constitutes a proof of concept that NPE-emulated RMF models, once validated, can be safely used for superfast exploratory inference. As an additional mock-observation test, imposing $R_{1.4}=12\,{\rm km}$ and $M_{\rm max}>1.97\,M_\odot$ leads to consistent predictions for the maximum-mass configuration, with DDB giving $M_{\rm max}=2.10^{+0.09}_{-0.07}\,M_\odot$, $R_{\rm max}=10.71^{+0.14}_{-0.21}\,{\rm km}$ and RMF-NL giving $M_{\rm max}=2.05^{+0.10}_{-0.06}\,M_\odot$, $R_{\rm max}=10.69^{+0.18}_{-0.19}\,{\rm km}$; although fixing $R_{1.4}$ confines both families to a narrow EOS region, RMF-NL remains marginally softer than DDB at high density, consistent with its slightly lower maximum mass.
\end{abstract}


\end{frontmatter}

\section{Introduction}
\label{introduction}
The equation of state (EoS) of dense matter is constrained by combining nuclear-theory inputs with a growing body of multimessenger data~\cite{Lattimer:2006xb,Burgio:2021vgk}. Gravitational-wave event such as GW170817~\cite{LIGOScientific:2017vwq} constrain the tidal deformability, NICER observations of the millisecond pulsars PSR~J0030+0451~\cite{Riley:2019yda,Miller:2019cac}, PSR~J0740+6620~\cite{Riley:2021pdl,Miller:2021qha}, PSR~J0437-4715~\cite{Choudhury:2024xbk}, and PSR~J1231-1411~\cite{Salmi:2024bss} provide independent mass--radius information, and chiral effective field theory (chiral EFT) fixes the low-density behavior of neutron-rich matter~\cite{Hebeler:2013nza,Tews:2012fj}. Complementing this from above, perturbative QCD (pQCD) constrains the EoS at asymptotically high densities~\cite{Kurkela:2009gj}. Although pQCD is reliable only near $n \sim 40\,n_{\rm sat}$, far above neutron-star cores, the requirements of thermodynamic stability and causality propagate this high-density input down to neutron-star densities and exclude a sizable part of the otherwise-allowed EoS band~\cite{Komoltsev:2021jzg}. Together, the chiral-EFT and pQCD limits already bracket a meaningful domain for the dense-matter EoS~\cite{Annala:2019puf}.

The volume and precision of these data will rise sharply with third-generation detectors such as the Einstein Telescope~\cite{ET:2025xjr} and Cosmic Explorer~\cite{Reitze:2019iox}. This progress exposes a computational bottleneck: Bayesian EoS inference~\cite{Steiner:2010fz,Raaijmakers:2019dks,Raaijmakers:2021uju,Huth:2021bsp,Annala:2019puf,Landry:2020vaw} requires posterior sampling over the model parameters together with repeated solutions of the stellar-structure equations, and the full analysis must be redone from scratch for every new measurement, revised uncertainty, or alternative EoS family. As the number of observations grows, the repeated computational cost of reanalyzing the data increases. To enable near-real-time inference in future detector analysis pipelines, alternative approaches to conventional Bayesian inference methods will be required. 

These limitations motivate simulation-based inference (SBI)~\cite{doi:10.1073/pnas.1912789117,BoeltsDeistler_sbi_2025}, in which synthetic parameter- observable pairs are drawn from the model across a wide prior and a neural density estimator is trained to approximate the posterior directly. Its defining advantage is amortization: the simulation and training cost is paid only once, after which the trained estimator is conditioned on new or higher-precision data in seconds, without resampling or retraining---precisely the regime that conventional likelihood-based inference, which must be repeated in full for every updated measurement, cannot reach. This advantage carries a known risk, however: amortized neural posteriors can be overconfident and statistically unfaithful~\cite{hermans2022trustcrisissimulationbasedinference}, so an SBI posterior cannot be trusted in a production analysis until it has been benchmarked against a reference Bayesian calculation and shown to be well calibrated.

In gravitational-wave astronomy this validation is well advanced---neural posterior estimators now reproduce standard-sampler results for compact-binary source parameters, from early demonstrations~\cite{Chua:2019wwt} to the DINGO framework and its real-time extensions~\cite{Dax:2021tsq,Dax:2022pxd,Dax:2024mcn}. For the dense-matter EoS, by contrast, SBI has so far been applied only to agnostic EoS representations~\cite{Carvalho:2025abc} or to neural likelihood estimation from X-ray spectra~\cite{Brandes:2024abc}, and in neither case has the amortized posterior been compared one-to-one with a nested-sampling analysis of the same physical model. Whether SBI can reproduce a full Bayesian analysis at the level of the microscopic parameters of a physical EoS model---the prerequisite for trusting amortized EoS inference in future detector pipelines---therefore remains untested. We address this gap directly, applying amortized neural posterior estimation to the couplings of relativistic mean-field (RMF) models for the first time and benchmarking it one-to-one against a nested-sampling pipeline, so that the resulting estimator can be reused as new neutron-star observations accumulate without rerunning a likelihood-based scan.

Neural posterior estimation is applied to two established RMF families---the density-dependent DDB model~\cite{malik2022} and the nonlinear RMF-NL model~\cite{Cartaxo:2025jpi}---using the standard minimal constraint set: nuclear saturation properties, chiral-EFT pure-neutron-matter pressures, and the $2\,M_\odot$ maximum-mass bound. Because the inference targets the RMF couplings themselves, the results are physically interpretable rather than expressed only as macroscopic EoS bands. To validate the approach, we compare the obtained SBI posterior with the one obtained in the conventional Bayesian inference through  PyMultiNest with identical data considered. We also explore its statistical reliability using the  Tests of Accuracy with Random Points (TARP) coverage diagnostic. 

The paper is organized as follows. Section~\ref{methodolgy} describes the models, training data, and NPE setup; Section~\ref{results} presents the inferred posteriors, their comparison with PyMultiNest, and the amortization test; Section~\ref{Summary} summarizes our conclusions.

\section{Methodology}
\label{methodolgy}
\subsection{The Models}
We have employed two different types of phenomenological equation of state.  Both equations of state are built within the relativistic mean-field description, in which nucleons interact through the exchange of the scalar-isoscalar $\sigma$, vector-isoscalar $\omega$, and vector-isovector $\rho$ mesons. A single Lagrangian density fixes the nuclear-matter energy functional and, through the meson field equations, the charge-neutral $\beta$-equilibrium EoS that is integrated with the Tolman--Oppenheimer--Volkoff (TOV) equations to yield the neutron-star observables. The two families differ only in how the interaction acquires its density dependence; their full Lagrangians, calibration, and parameter tables are given in Refs.~\cite{malik2022,Cartaxo:2025jpi}, and both are evaluated with the CompactObject package~\cite{Cartaxo:2025jpi}.
 
In the density-dependent model (DDB), the meson--nucleon couplings run with baryon density, $\Gamma_i(\rho)=\Gamma_{i,0}\,h_i(\rho/\rho_0)$ for $i=\sigma,\omega,\rho$, where the isoscalar functions are controlled by the exponents $a_\sigma,a_\omega$ and the isovector coupling decreases as $h_\rho(x)=\exp[-a_\rho(x-1)]$~\cite{malik2022,Cartaxo:2025jpi}. The six inferred parameters are therefore the saturation-density couplings $\Gamma_{\sigma,0}$, $\Gamma_{\omega,0}$, $\Gamma_{\rho,0}$ together with the density-slope parameters $a_\sigma$, $a_\omega$, $a_\rho$.
 
In the nonlinear model (RMF-NL), the couplings $\Gamma_\sigma$, $\Gamma_\omega$, $\Gamma_\rho$ are constant and the density dependence is generated instead by nonlinear meson self-interactions: cubic and quartic $\sigma$ terms with strengths $\kappa$ and $\lambda_0$, a quartic $\omega$ term $\zeta$, and an $\omega$--$\rho$ cross-coupling $\Lambda_\omega$~\cite{Cartaxo:2025jpi}. These terms carry distinct physical roles---$\kappa$ and $\lambda_0$ set the saturation incompressibility, $\zeta$ softens the high-density EoS, and $\Lambda_\omega$ controls the density dependence of the symmetry energy---giving the seven inferred parameters $(\Gamma_\sigma,\Gamma_\omega,\Gamma_\rho,\kappa,\lambda_0,\zeta,\Lambda_\omega)$. Prior ranges for both of the model parameters considered in this study are shown in Table~\ref{tab:ddb_rmfnl_training_bank} 
 
Both families are calibrated against the same minimal set of constraints used throughout this work---nuclear saturation properties, chiral-EFT pure-neutron-matter pressures, and the $2\,M_\odot$ maximum-mass bound (Table~\ref{tab:fiducial_input})---so that the simulation-based and nested-sampling posteriors are compared on an identical footing. We treat both EoS families as fixed forward models and focus on the inference methodology.

\subsection{Simulation-Based Inference}
\label{sec:sbi_method}

We employ neural posterior estimation (NPE)~\cite{papamakarios2018fastepsilonfreeinferencesimulation}, implemented using the sbi package~\cite{BoeltsDeistler_sbi_2025}, to infer the coupling parameters of the DDB and RMF-NL models from nuclear-matter and neutron-star observables. For each model, a large prior-predictive bank of parameter--observable pairs, $(\boldsymbol{\theta},\mathbf{x})$, is generated using the corresponding RMF and stellar-structure forward pipeline. A conditional neural spline flow is then trained to approximate
\begin{equation}
p(\boldsymbol{\theta}\mid\mathbf{x},\boldsymbol{\sigma})
\simeq
q_{\phi}(\boldsymbol{\theta}\mid\mathbf{x},\boldsymbol{\sigma}),
\end{equation}
where $\boldsymbol{\sigma}$ denotes the observational-uncertainty vector supplied as an additional conditioning input. During training, the uncertainty scale is varied continuously over the range $0.5$--$2$ times the reference uncertainties, allowing a single amortized estimator to be applied at different observational precisions. Posterior calibration is assessed on held-out simulations using the TARP diagnostic~\cite{2023PMLR..20219256L}; where required, coverage and residual location corrections are applied before imposing the astrophysical condition $M_{\max}>2\,M_{\odot}$. The resulting posterior samples are propagated through the forward model to obtain the EoS and derived neutron-star observables, which are compared with independent nested-sampling results. Complete details of the training-bank construction, network architecture, optimization, validation, calibration, computational cost, and posterior-processing procedure are
provided in the Supplemental Material.

\begin{table}[t]
\centering
\caption{Prior and prior-predictive ranges used to construct the representative DDB and RMF-NL training banks.}
\label{tab:ddb_rmfnl_training_bank}
\scriptsize
\renewcommand{\arraystretch}{1.1}
\setlength{\tabcolsep}{3pt}
\begin{tabular*}{\columnwidth}{@{\extracolsep{\fill}}l c c | l c c@{}}
\hline
\multicolumn{3}{c|}{\textbf{DDB}} & \multicolumn{3}{c}{\textbf{RMF-NL}} \\
\hline
Quantity & Min. & Max. & Quantity & Min. & Max. \\
\hline
\multicolumn{3}{l|}{\textit{Prior}} & \multicolumn{3}{l}{\textit{Prior}} \\
$a_\sigma$ & 0 & 0.300 & $\Gamma_{\sigma}$ & 6.50 & 15.50 \\
$a_\omega$ & 0 & 0.298 & $\Gamma_{\omega}$ & 6.50 & 15.50 \\
$a_\rho$ & 0 & 1.300 & $\Gamma_{\rho}$ & 5.50 & 16.50 \\
$\Gamma_{\sigma,0}$ & 6.50 & 13.49 & $\kappa$ & 0.0048 & 0.086 \\
$\Gamma_{\omega,0}$ & 7.50 & 14.50 & $\lambda_0$ & -0.05 & 0.05 \\
$\Gamma_{\rho,0}$ & 5.00 & 12.55 & $\zeta$ & 0.0 & 0.04 \\
\multicolumn{3}{l|}{} & $\Lambda_\omega$ & 0.0 & 0.15 \\
\hline
\multicolumn{3}{l|}{\textit{Outputs}} & \multicolumn{3}{l}{\textit{Outputs}} \\
$\rho_0$ [fm$^{-3}$] & 0.140 & 0.170 & $\rho_0$ [fm$^{-3}$] & 0.140 & 0.170 \\
$E_0$ [MeV] & -100 & 100 & $E_0$ [MeV] & -99.96 & 100.0 \\
$K_0$ [MeV] & $\sim0$ & 800 & $K_0$ [MeV] & $\sim0$ & 800.0 \\
$J_{\rm sym}$ [MeV] & 19.2 & 50.0 & $J_{\rm sym}$ [MeV] & 16.58 & 50.0 \\
$P_{\rm PNM}(0.08)$ & $\sim0$ & 7.60 & $P_{\rm PNM}(0.08)$ & 0.0 & 7.71 \\
$P_{\rm PNM}(0.12)$ & $\sim0$ & 18.45 & $P_{\rm PNM}(0.12)$ & 0.41 & 17.95 \\
$P_{\rm PNM}(0.16)$ & $\sim0$ & 34.67 & $P_{\rm PNM}(0.16)$ & 1.25 & 33.19 \\
$M_{\rm max}$ [$M_\odot$] & 1.40 & 3.56 & $M_{\rm max}$ [$M_\odot$] & 1.40 & 3.49 \\
\hline
\end{tabular*}
\end{table}

\begin{figure*}[t]
    \centering

    \begin{minipage}[t]{0.99\textwidth}
        \centering
        \includegraphics[width=\linewidth]{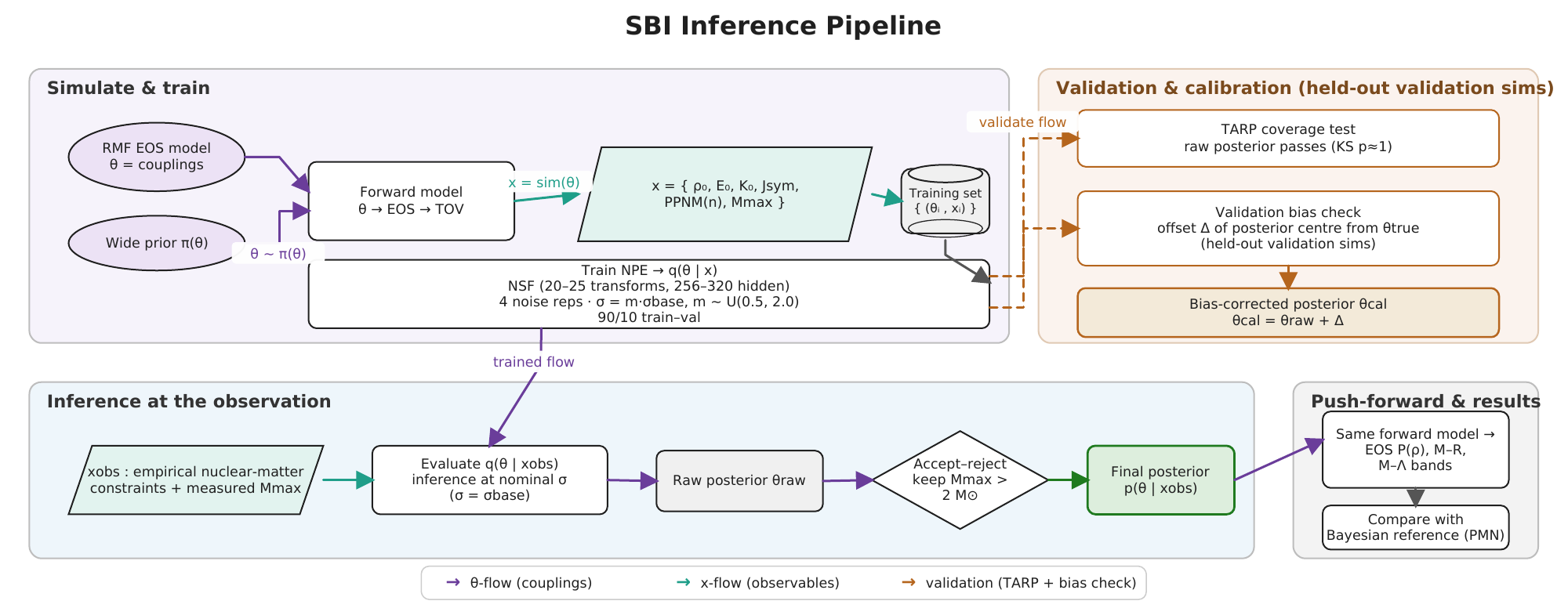}
    \end{minipage}

     \caption{Schematic of the SBI training pipeline.}
    \label{flowchart}
\end{figure*}

\begin{table}[t]
\centering

\caption{
Nuclear-matter and astrophysical inputs used in this study. For each quantity written as \(\mu\pm\sigma_{\rm base}\), \(\mu\) denotes the central value used in the fiducial observation vector \(\mathbf{x}_{\rm obs}\), while \(\sigma_{\rm base}\) denotes the corresponding \(1\sigma\) uncertainty. During NPE training, noise-augmented replicas are generated using scaled uncertainties \(\boldsymbol{\sigma}=m\,\boldsymbol{\sigma}_{\rm base}\), with \(m\sim\mathcal{U}(0.5,2.0)\). The quantities \(E_0\), \(K_0\), and \(J_{\rm sym}\) are given in MeV, \(P_{\rm PNM}\) in MeV fm\(^{-3}\), \(\rho_0\) in fm\(^{-3}\), and \(M_{\rm max}\) in \(M_\odot\). The PNM pressure uncertainties correspond to \(2\times{\rm N}^3{\rm LO}\) chiral-EFT errors. }

\label{tab:fiducial_input}
\footnotesize
\renewcommand{\arraystretch}{1.18}
\setlength{\tabcolsep}{3.5pt}
\begin{tabular}{@{}p{0.27\columnwidth}p{0.25\columnwidth}p{0.38\columnwidth}@{}}
\hline
Quantity & Input & Source / use \\
\hline
$E_0$ 
& $-16.1 \pm 0.2$ 
& Dutra et al. (2014)~\cite{Dutra:2014qga} \\

$K_0$ 
& $230 \pm 40$ 
& Todd-Rutel and Piekarewicz (2005); Shlomo et al. (2006)~\cite{Shlomo:2006ole} \\

$J_{\rm sym}$ 
& $32.5 \pm 1.8$ 
& Essick et al. (2021)~\cite{Essick:2021kjb}\\

$P_{\rm PNM}(0.08)$ 
& $0.505714 \pm 0.194286$ 
& Hebeler et al. (2013)~\cite{Hebeler:2013nza} \\

$P_{\rm PNM}(0.12)$ 
& $1.241429 \pm 0.608571$ 
& \\

$P_{\rm PNM}(0.16)$ 
& $2.485714 \pm 1.382857$ 
& \\

$\rho_0$ 
& $0.153 \pm 0.005$ 
& Typel and Wolter (1999)~\cite{Typel:1999yq}\\

$M_{\rm max}$ 
& $>2.0$ 
& Fonseca et al. (2016)~\cite{Fonseca:2016tux} \\
\hline
\end{tabular}
\end{table}

\section{Results $\&$ Discussion}
\label{results}

\begin{figure*}[t]
    \centering

    \begin{minipage}[t]{0.32\textwidth}
        \centering
        \includegraphics[width=\linewidth]{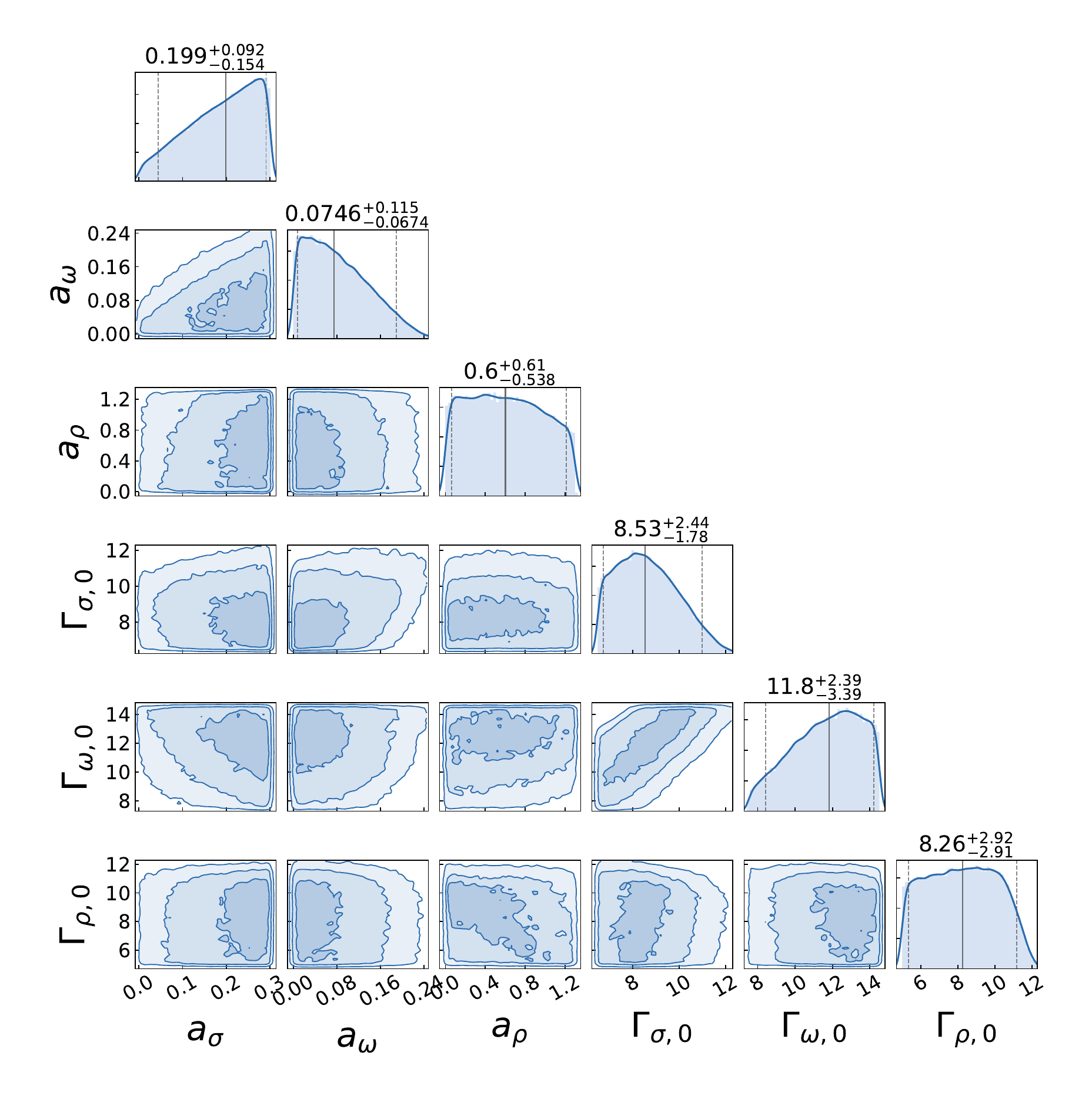}
    \end{minipage}
    \hfill
    \begin{minipage}[t]{0.32\textwidth}
        \centering
        \includegraphics[width=\linewidth]{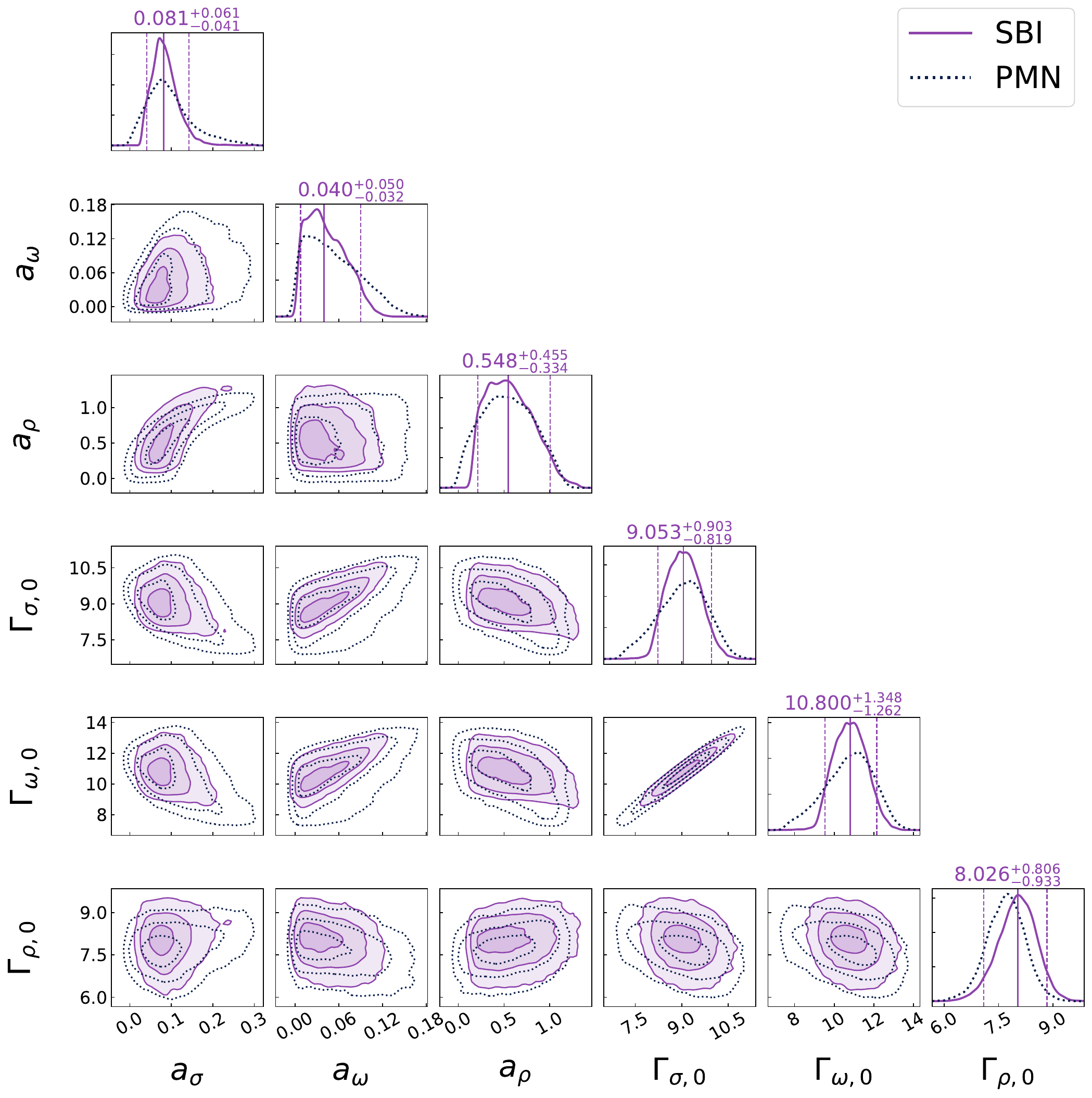}
    \end{minipage}
    \hfill
    \begin{minipage}[t]{0.32\textwidth}
        \centering
        \includegraphics[width=\linewidth]{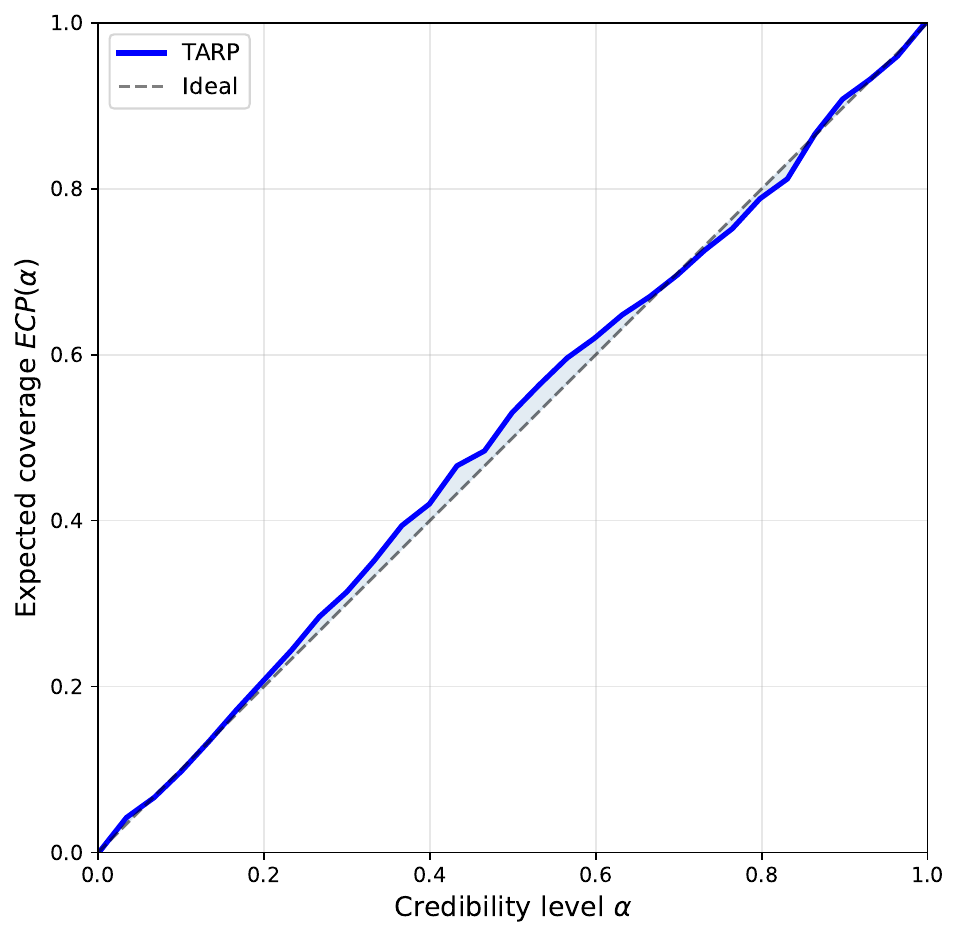}
    \end{minipage}

     \caption{Left: prior distributions of the DDB model parameters used in the inference. Middle: posterior distributions inferred with Simulation Based Inference (SBI), compared against the PyMultiNest (PMN) benchmark through one- and two-dimensional marginal contours. The numerical values above the one-dimensional KDE panels denote the median parameter, with the quoted lower and upper uncertainties corresponding to the 90\% credible interval. Right: Tests of Accuracy with Random Points (TARP) calibration diagnostic, where the expected coverage probability closely follows the ideal calibrated relation.}
    \label{fig:compare_tarp_ddb}
\end{figure*}

\begin{figure*}[t]
    \centering

    \begin{minipage}[t]{0.32\textwidth}
        \centering
        \includegraphics[width=\linewidth]{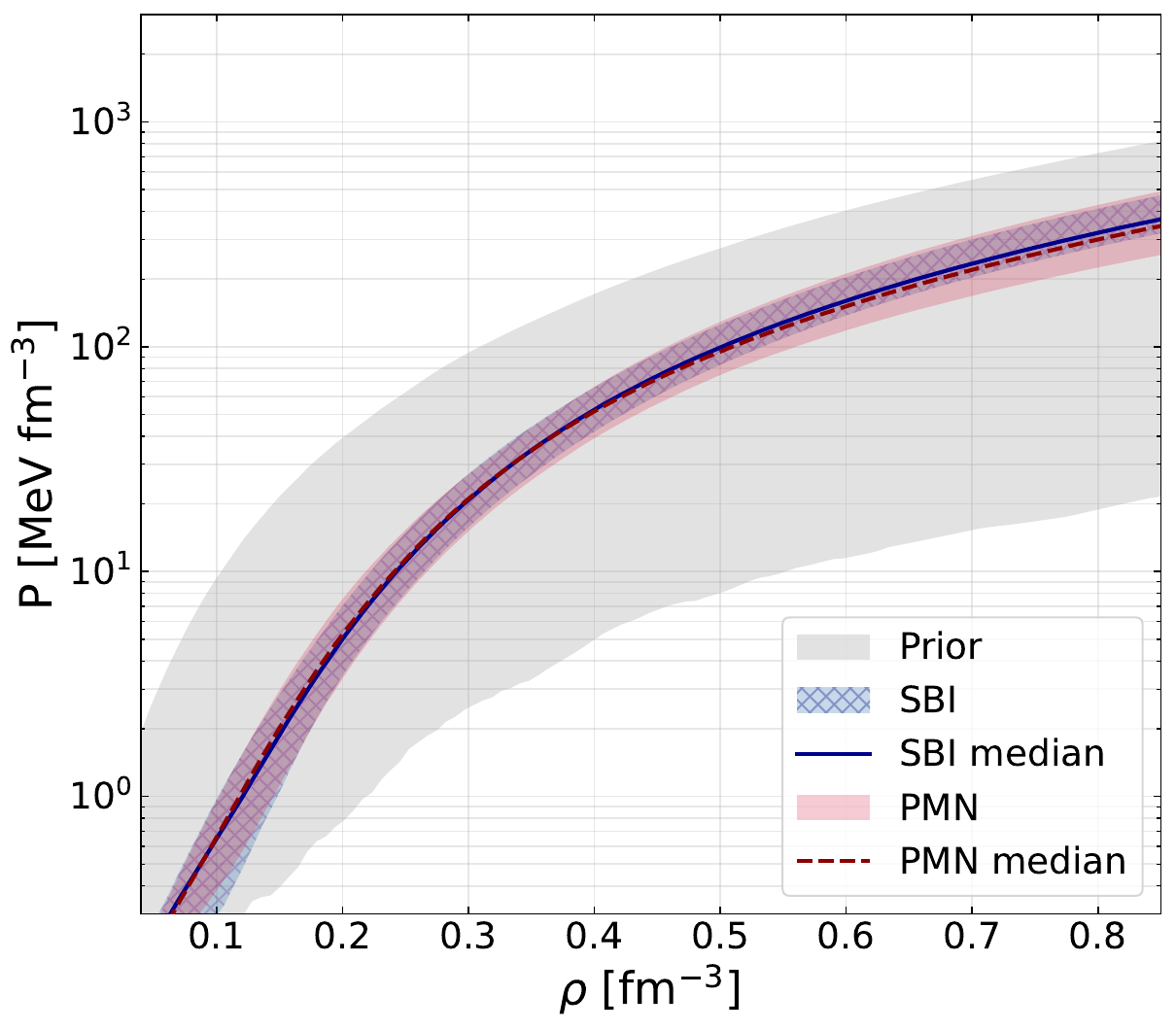}
    \end{minipage}
    \hfill
    \begin{minipage}[t]{0.32\textwidth}
        \centering
        \includegraphics[width=\linewidth]{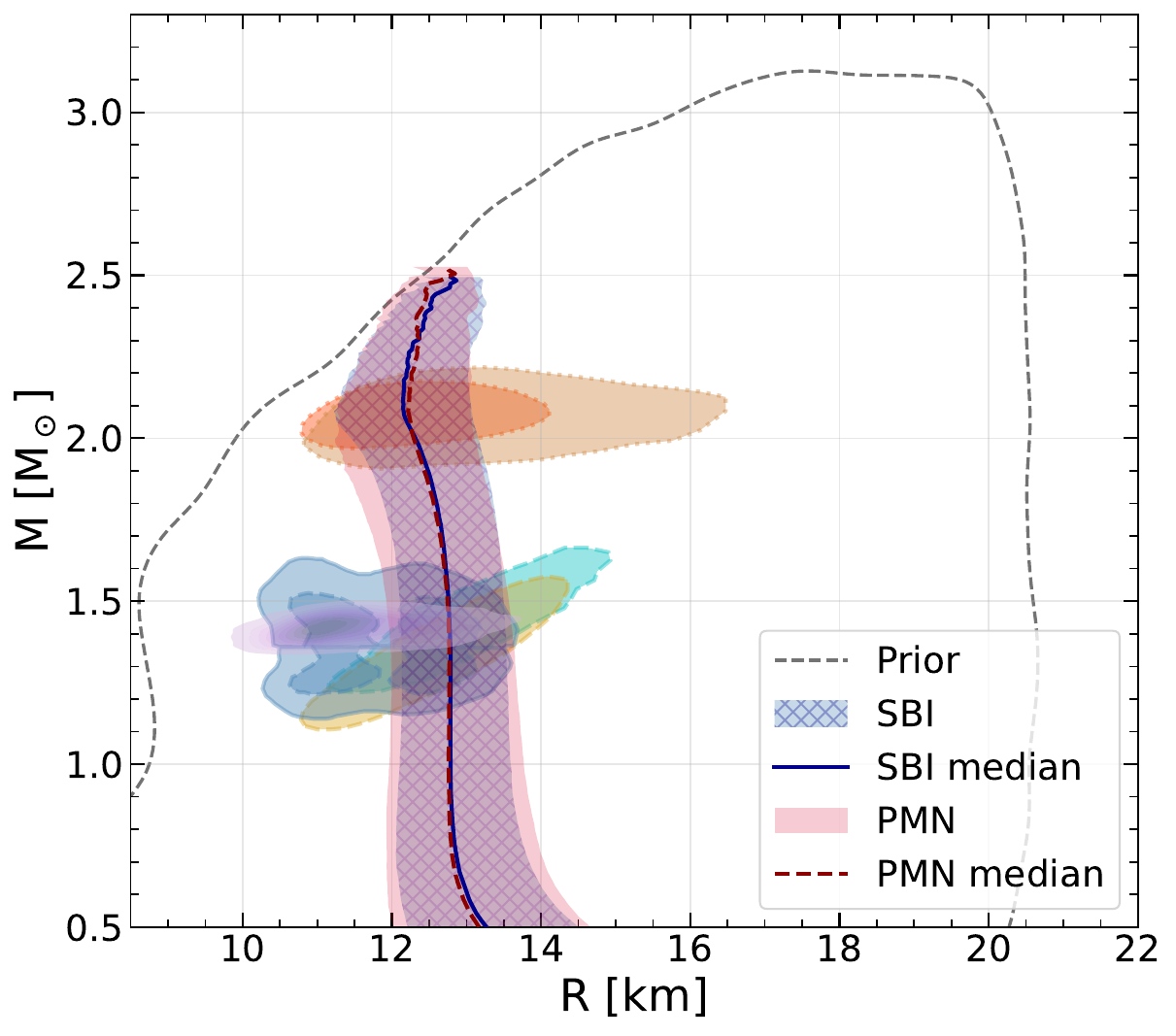}
    \end{minipage}
    \hfill
    \begin{minipage}[t]{0.32\textwidth}
        \centering
        \includegraphics[width=\linewidth]{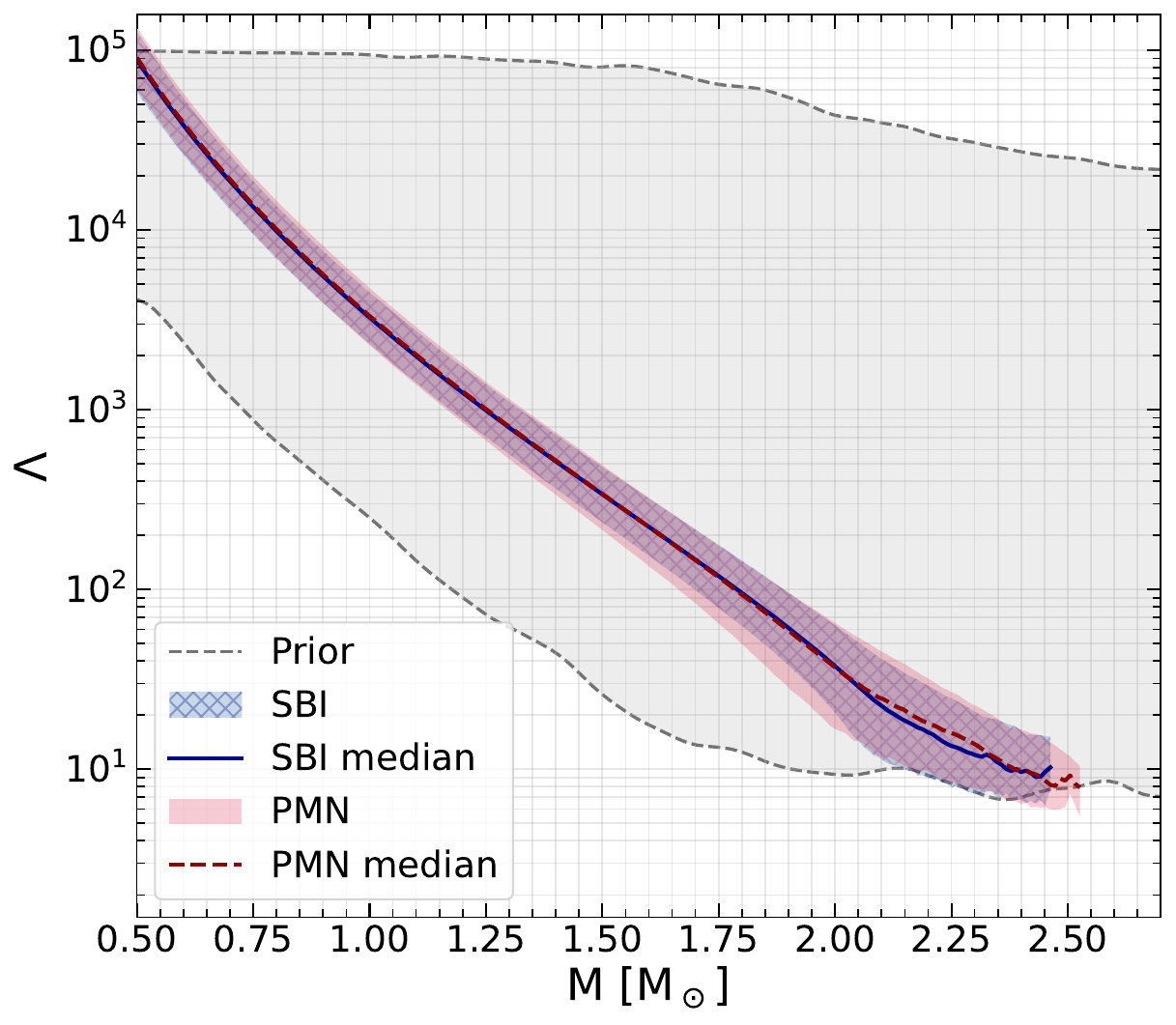}
    \end{minipage}

    \caption{Left: DDB posterior equation-of-state predictions for pressure $P$ as a function of baryon density $\rho$. Middle: mass--radius posterior regions, including the prior envelope and observational mass--radius constraints as follows: steel blue indicates the binary components of GW170817, with their respective 90\% and 50\% credible intervals. The plot includes the 1 $\sigma$ (68\%) CI for the 2D mass-radius posterior distributions of the millisecond pulsars PSR J0030 + 0451 (in cyan and yellow color) \cite{Riley:2019yda, Miller:2019cac} and PSR J0740 + 6620 (in orange and peru color)\cite{Riley:2021pdl, Miller:2021qha}. Furthermore, we display the latest NICER measurements for the mass and radius of PSR J0437-4715 \cite{Choudhury:2024xbk} (lilac color).Right: mass--tidal deformability relation $\Lambda(M)$. In all panels, the SBI results are compared with the PyMultiNest benchmark through the corresponding credible regions and median curves.}
    \label{fig:eos_mr_mtidal_compare_ddb}
\end{figure*}

Figure~\ref{fig:compare_tarp_ddb} presents the validation of the SBI posterior for the DDB parametrization through a direct comparison with the PyMultiNest benchmark and an independent statistical calibration test. The left panel shows the prior distributions adopted for the six DDB model parameters, demonstrating that the inference is initialized over broad parameter ranges before imposing the observational constraints. The middle panel compares the one- and two-dimensional marginal posterior distributions obtained from SBI and PyMultiNest. The agreement is very good: the SBI contours closely follow the PyMultiNest contours, and the dominant parameter correlations are consistently reproduced. For example, the strong positive correlation between $\Gamma_{\sigma,0}$ and $\Gamma_{\omega,0}$ is accurately captured by SBI, as shown by the similar elongated contour structure in the corresponding two-dimensional posterior plane. Similarly, the correlations involving $a_{\sigma}$, $a_{\omega}$, and $a_{\rho}$ are reproduced with the same overall orientation and spread as in the PyMultiNest benchmark. This visual agreement is supported quantitatively by the posterior summaries listed in Table~\ref{tab:ddrmf_rmfnl_combined}. For SBI, the inferred median values are $a_{\sigma}=0.081$, $a_{\omega}=0.040$, $a_{\rho}=0.548$, $\Gamma_{\sigma,0}=9.053$, $\Gamma_{\omega,0}=10.800$, and $\Gamma_{\rho,0}=8.026$, while the corresponding PyMultiNest medians are $0.087$, $0.047$, $0.534$, $9.068$, $10.835$, and $7.711$, respectively. The shifts in the median values are small compared with the corresponding $90\%$ credible intervals, and the uncertainty ranges overlap for all six parameters. This indicates that SBI does not introduce any significant bias relative to the nested-sampling reference. In some parameters, such as $a_{\sigma}$, $a_{\omega}$, $\Gamma_{\sigma,0}$, and $\Gamma_{\omega,0}$, the SBI posterior is slightly more compact than the PyMultiNest posterior; however, this difference remains within the statistical uncertainty and does not change the inferred physical parameter region.
The right panel further tests the reliability of the SBI posterior using the
Tests of Accuracy with Random Points (TARP) diagnostic~\cite{2023PMLR..20219256L}. The test uses
$N_{\rm sim}=500$ prior-predictive validation simulations and
$N_{\rm s}=1000$ posterior samples per validation. The TARP diagnostic was
evaluated using the built-in \texttt{run\_tarp} routine of the \texttt{sbi}
package~\cite{BoeltsDeistler_sbi_2025}. In this test,
prior-predictive validation pairs
$(\boldsymbol{\theta}_i,\mathbf{x}_i)$ are generated, posterior samples
$\boldsymbol{\theta}_{ij}\sim p(\boldsymbol{\theta}\mid \mathbf{x}_i)$
are drawn, and a random reference point
$\boldsymbol{\theta}_{\rm r}$ is chosen in the parameter space. For each
validation pair, one computes the fraction of posterior samples whose
distance from the reference point is smaller than the corresponding distance
of the true parameter, namely
\begin{equation}
r_i =
\frac{1}{N_{\rm s}}
\sum_{j=1}^{N_{\rm s}}
\mathbb{I}
\left[
d\!\left(\boldsymbol{\theta}_{ij},\boldsymbol{\theta}_{\rm r}\right)
<
d\!\left(\boldsymbol{\theta}_{i},\boldsymbol{\theta}_{\rm r}\right)
\right],
\end{equation}
where $N_{\rm s}$ is the number of posterior samples and $\mathbb{I}$
denotes the indicator function. The TARP expected coverage probability is
then obtained by evaluating the empirical coverage over many validation
simulations as a function of the credibility level $\alpha$. For a calibrated
posterior, the expected relation is
\begin{equation}
\mathrm{ECP}(\alpha)=\alpha,
\end{equation}
which corresponds to the diagonal ideal-calibration line. The TARP curve in
the right panel closely follows this ideal relation over the full credibility
range, with only small statistical fluctuations. Therefore, the SBI posterior
does not show systematic overconfidence, which would appear as undercoverage,
or excessive underconfidence, which would appear as overcoverage. Taken
together, the close agreement with PyMultiNest and the successful TARP
calibration test demonstrate that SBI provides an accurate, statistically
calibrated, and computationally efficient approximation to the full Bayesian
posterior for the DDB model.

Left panel of figure~\ref{fig:eos_mr_mtidal_compare_ddb} shows the propagation of the inferred DDB posterior to the EoS and NS observables, again comparing SBI with the PyMultiNest benchmark. In the left panel, the posterior pressure-density relation, $P(\rho)$, is significantly more constrained than the prior envelope, showing that the observational data strongly reduce the allowed EoS uncertainty. The SBI and PMN credible bands almost completely overlap, and their median curves follow each other over the full density range, indicating that both methods infer nearly the same pressure support for the DDB model. At the $90\%$ credible level, the only appreciable difference between the two posteriors appears in the high-density EoS, where the DDB band inferred by SBI is marginally stiffer than the PMN one. This region is fixed almost entirely by the maximum-mass condition rather than by the saturation and chiral-EFT inputs, and is therefore the least directly constrained part of the EoS; moreover, because the stiffer DDB EoS reaches its maximum mass at a slightly lower central density than the softer RMF-NL EoS, its highest densities are probed by comparatively few stellar configurations. Although several factors may contribute, we believe this limited high-density sampling is one plausible reason for the small residual, which in any case lies within the $90\%$ credible interval and does not affect the inferred neutron-star observables.
 The consistency between SBI and PMN is also evident in the derived nuclear matter properties listed in Table~\ref{tab:ddrmf_rmfnl_combined}. Using the $90\%$ credible-interval width, we find comparable uncertainty estimates from the two methods. For $K_0$, $J_{\rm sym}$, $L_{\rm sym}$, and $K_{\rm sym}$, the SBI widths are $102.5$, $6.13$, $87.2$, and $147.1~{\rm MeV}$, while the corresponding PMN widths are $110.4$, $5.63$, $100.0$, and $150.0~{\rm MeV}$. The middle panel shows the posterior mass--radius relation obtained after propagating the DDB posterior to stellar structure. Compared with the broad prior envelope, the posterior region is strongly reduced and lies within the observationally allowed mass--radius domain. The SBI and PMN credible regions overlap closely, and their median curves are nearly indistinguishable along the stable branch. This agreement is also seen from the $90\%$ credible-interval widths in Table~\ref{tab:ddrmf_rmfnl_combined}: SBI gives $\Delta_{90}(M_{\rm max})=0.279\,M_{\odot}$, $\Delta_{90}(R_{\rm max})=1.18\,{\rm km}$, and $\Delta_{90}(R_{1.4})=1.34\,{\rm km}$, compared with the PMN values $0.476\,M_{\odot}$, $1.40\,{\rm km}$, and $1.65\,{\rm km}$, respectively. The right panel shows the corresponding mass--tidal deformability relation, $\Lambda(M)$. Both SBI and PMN predict the same monotonic decrease of $\Lambda$ with increasing stellar mass, and the posterior bands overlap over the full mass range. At the canonical mass, the median values differ by only about $1\%$, with $\Lambda_{1.4}=516.0$ from SBI and $\Lambda_{1.4}=521.1$ from PMN. The $90\%$ credible-interval widths are also comparable, $\Delta_{90}(\Lambda_{1.4})=361.5$ for SBI and $409.7$ for PMN. Overall, the comparison of the $90\%$ credible-interval widths shows that SBI reproduces the PMN uncertainty estimates for the nuclear matter properties, mass--radius relation, and tidal deformability with comparable credible-region sizes. The SBI posterior gives slightly more compact intervals for several quantities, including $K_0$, $L_{\rm sym}$, $K_{\rm sym}$, $M_{\rm max}$, $R_{1.4}$, and $\Lambda_{1.4}$, but these differences remain statistically consistent with the PMN benchmark. Thus, SBI accurately reproduces the PMN posterior predictions without introducing a significant bias in either the inferred EoS properties or the resulting neutron-star observables.

\begin{figure*}[t]
    \centering

    \begin{minipage}[t]{0.32\textwidth}
        \centering
        \includegraphics[width=\linewidth]{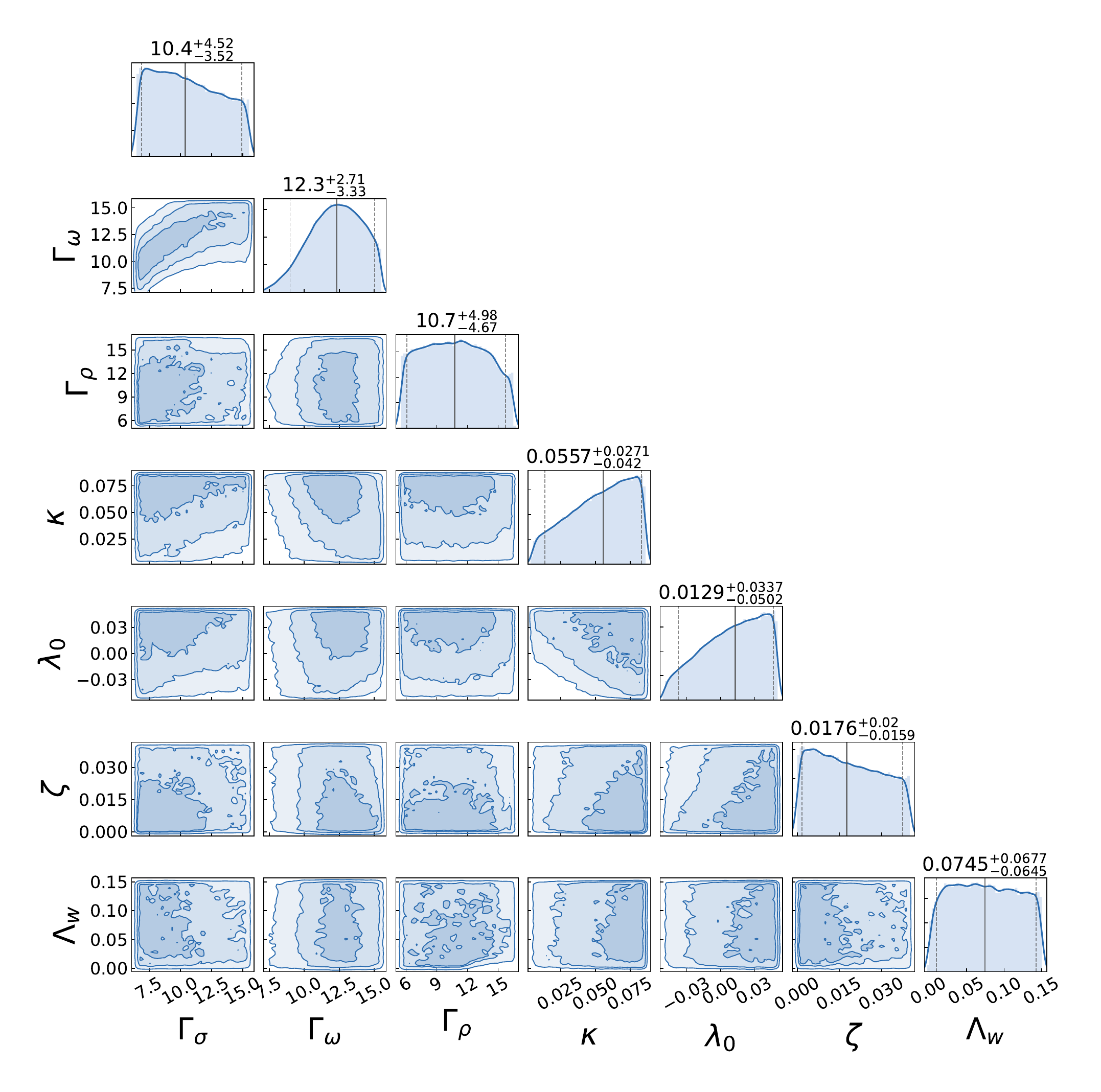}
    \end{minipage}
    \hfill
    \begin{minipage}[t]{0.32\textwidth}
        \centering
        \includegraphics[width=\linewidth]{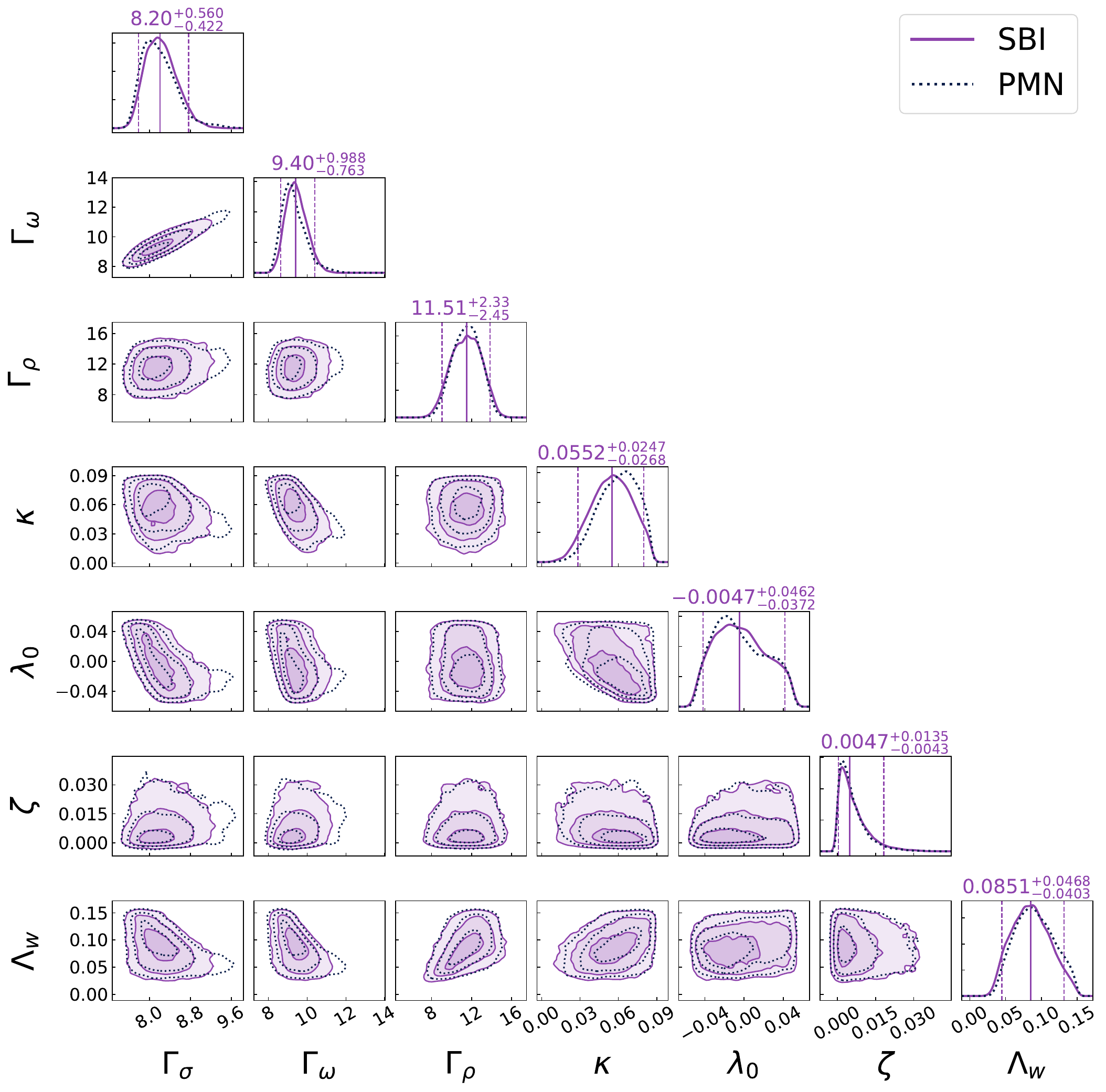}
    \end{minipage}
    \hfill
    \begin{minipage}[t]{0.32\textwidth}
        \centering
        \includegraphics[width=\linewidth]{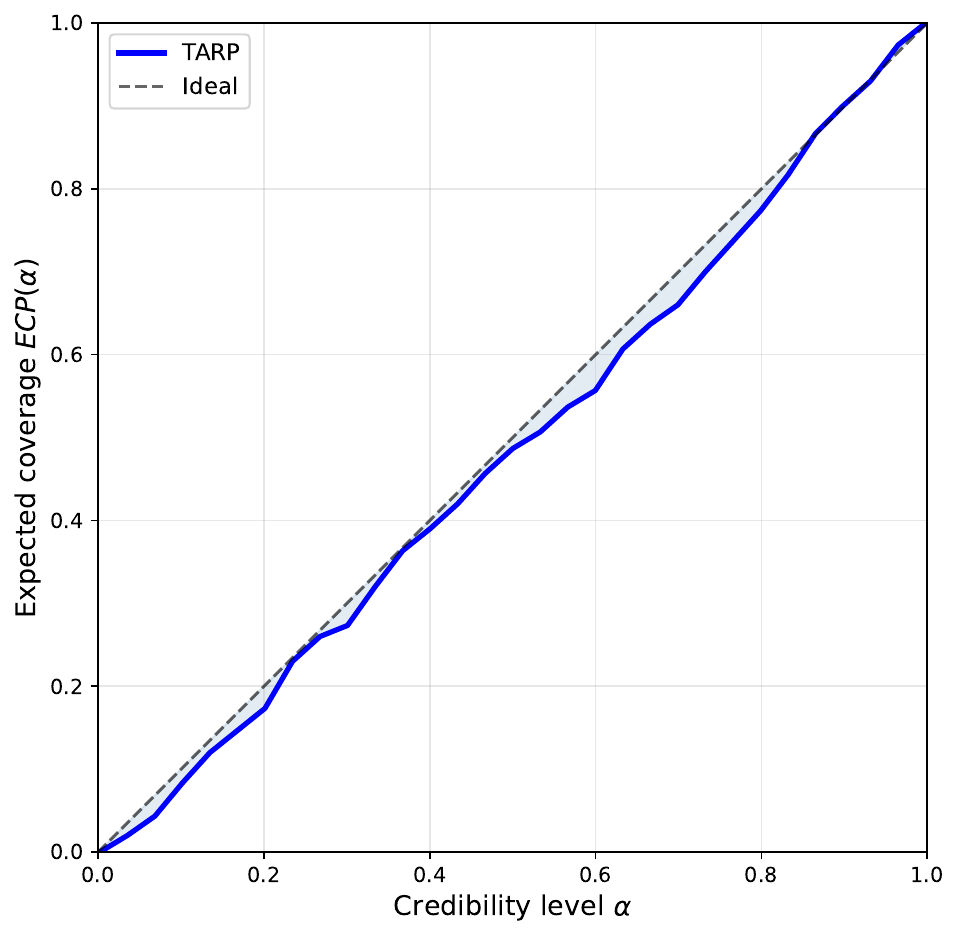}
    \end{minipage}

     \caption{
Left: prior distributions of the RMF-NL coupling parameters. 
Middle: SBI posterior distributions compared with the PyMultiNest benchmark, with the quoted values denoting median estimates and 90\% credible intervals. 
Right: TARP calibration test showing the expected coverage probability against the ideal calibrated relation.}
    \label{fig:compare_tarp_rmfnl}
\end{figure*}

\begin{figure*}[t]
    \centering

    \begin{minipage}[t]{0.32\textwidth}
        \centering
        \includegraphics[width=\linewidth]{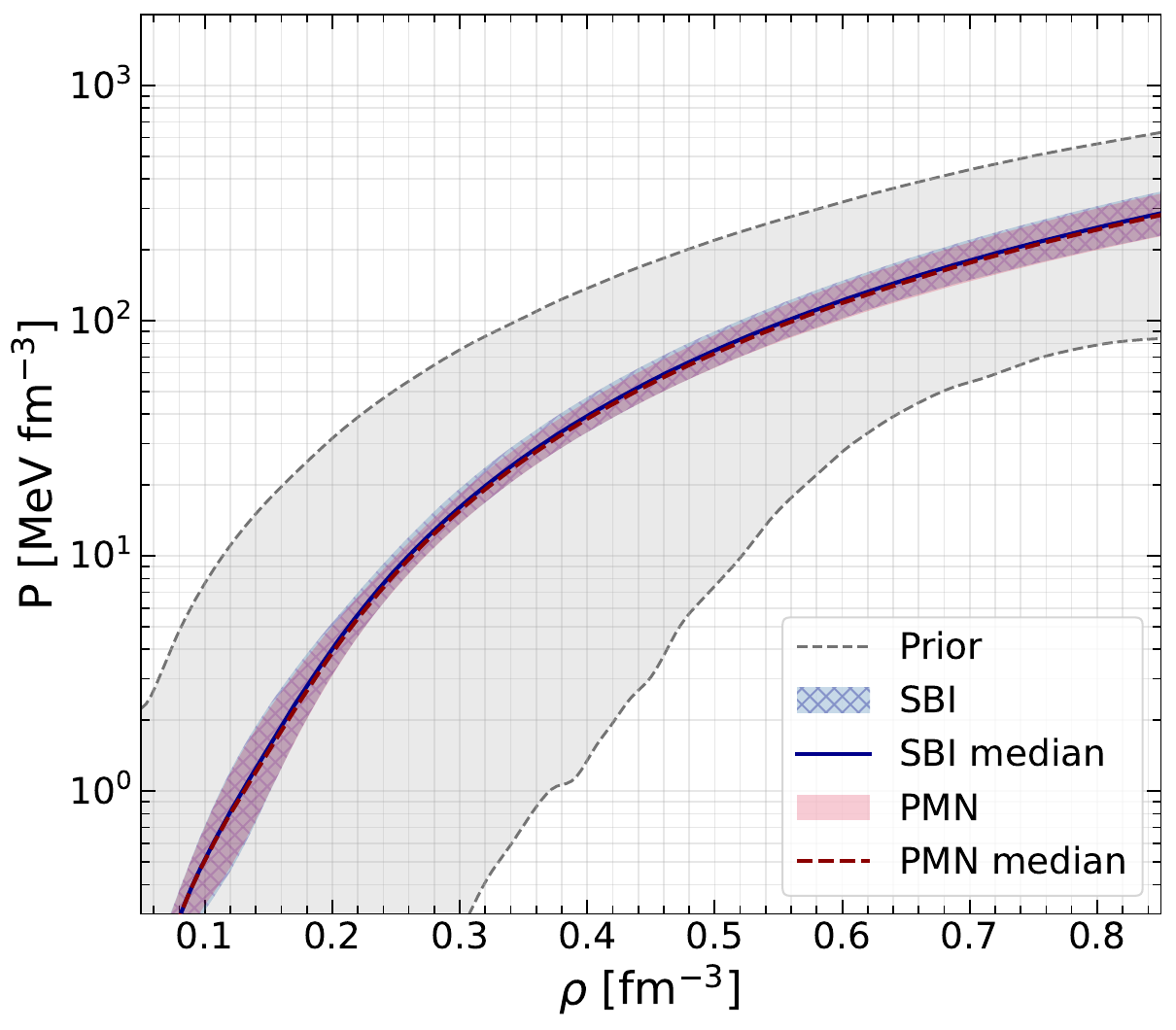}
    \end{minipage}
    \hfill
    \begin{minipage}[t]{0.32\textwidth}
        \centering
        \includegraphics[width=\linewidth]{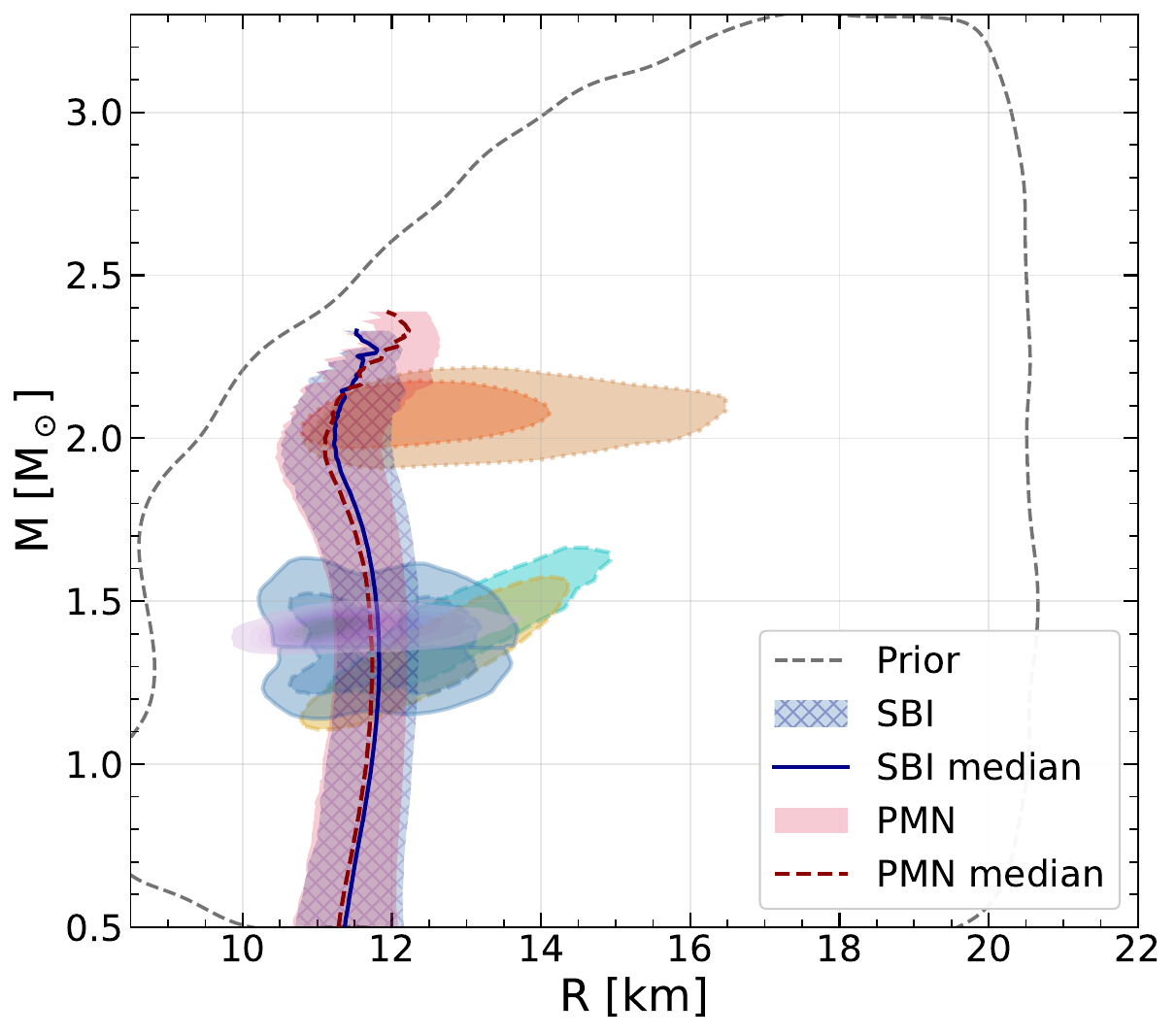}
    \end{minipage}
    \hfill
    \begin{minipage}[t]{0.32\textwidth}
        \centering
        \includegraphics[width=\linewidth]{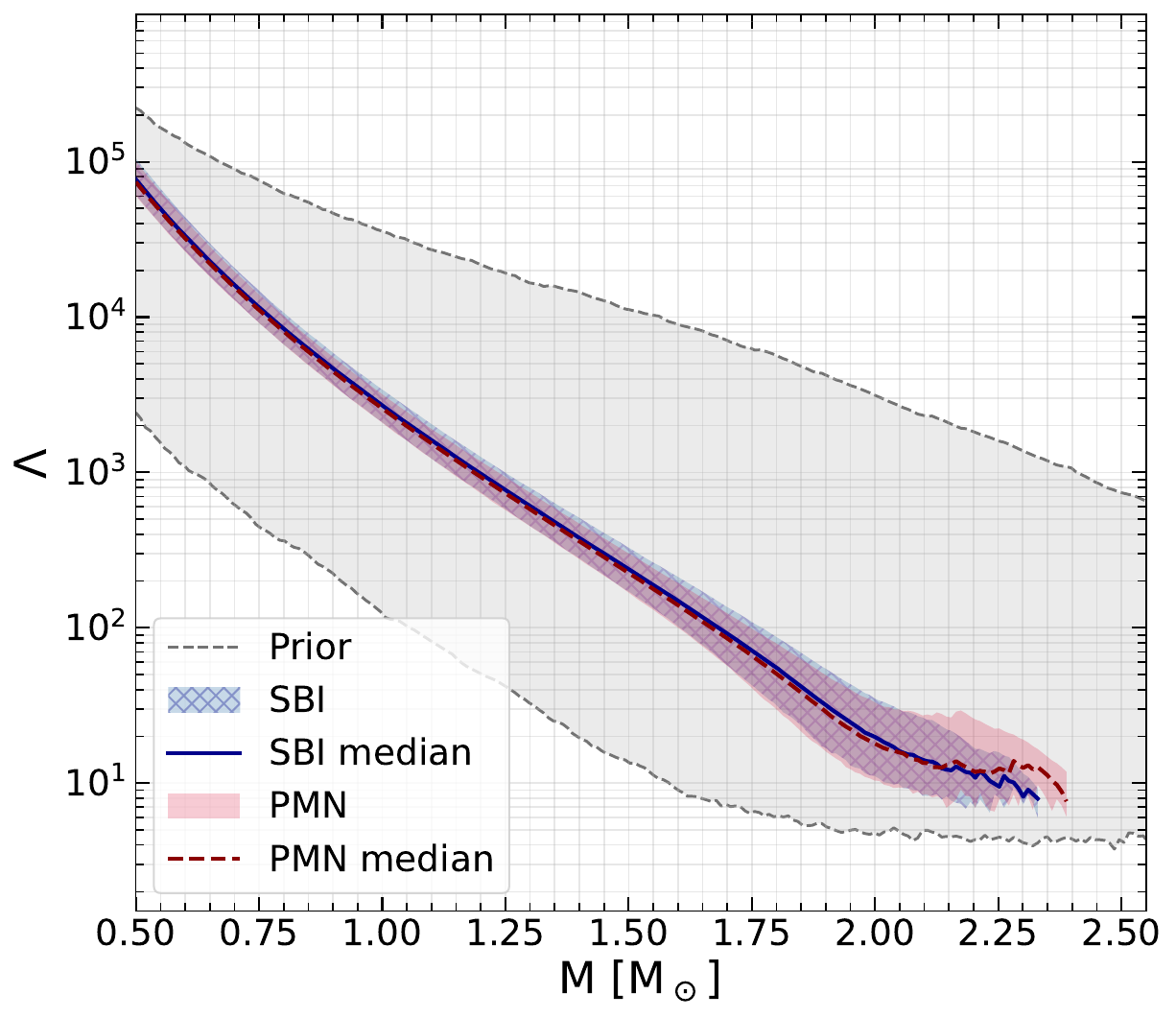}
    \end{minipage}

    \caption{Left: posterior RMF-NL equation-of-state predictions for pressure $P$ as a function of baryon density $\rho$. 
Middle: corresponding mass--radius posterior regions, together with the prior envelope and observational mass--radius constraints. 
Right: predicted mass--tidal deformability relation $\Lambda(M)$. 
In all panels, the SBI results are compared with the PyMultiNest benchmark using the corresponding median trends and credible regions.}
    \label{fig:eos_mr_mtidal_compare_rmfnl}
\end{figure*}

For the RMF-NL parametrization, the SBI posterior shows the same level of agreement with the PMN benchmark as obtained for the DDB case, as shown in figure~\ref{fig:compare_tarp_rmfnl}. The one-dimensional marginals and two-dimensional posterior contours closely overlap, indicating that SBI captures both the individual parameter constraints and the dominant correlations among the RMF-NL couplings. This is particularly visible in the correlated structures involving $\Gamma_\sigma$, $\Gamma_\omega$, and $\kappa$, where the SBI contours reproduce the orientation and spread of the PMN posterior. In the right panel of Fig.~\ref{fig:compare_tarp_rmfnl}, the corresponding TARP diagnostic remains close to the ideal calibration line, with only mild deviations at intermediate credibility levels.  For the RMF-NL posterior, a mild temperature rescaling with (T=1.20) improves the TARP calibration. The empirical coverage increases from (62.3\%) to (64.7\%) at the nominal (68\%) credibility level, and from (87.4\%) to (90.1\%) at the nominal (90\%) level. This reduces the mild intermediate-level undercoverage and restores the high-credibility coverage, indicating no systematic coverage bias in the calibrated RMF-NL SBI posterior. The propagated predictions for the RMF-NL EoS, mass--radius relation, and tidal deformability are shown in Fig.~\ref{fig:eos_mr_mtidal_compare_rmfnl}. The SBI and PMN pressure--density bands almost fully overlap, and their median EoS curves remain very close over the full density range. The mass--radius posterior is also consistently reproduced, with both methods giving compact stellar radii around $R_{1.4}\simeq 11.8~{\rm km}$ and maximum masses close to $2.0\,M_\odot$. The $\Lambda(M)$ relation shows the same behavior: the SBI and PMN bands overlap over the full mass range, and the canonical tidal deformability differs only mildly, with $\Lambda_{1.4}=378.0$ for SBI and $\Lambda_{1.4}=360.5$ for PMN.

The uncertainty comparison in Table~\ref{tab:ddrmf_rmfnl_combined} further confirms this consistency. Using the $90\%$ credible-interval width, the RMF-NL nuclear matter widths for $(K_0,J_{\rm sym},L_{\rm sym},K_{\rm sym})$ are $(85.7,5.59,37.2,101.0)~{\rm MeV}$ for SBI and $(73.2,5.07,34.8,105.7)~{\rm MeV}$ for PMN. For the macroscopic observables, SBI gives $\Delta_{90}(M_{\rm max})=0.315\,M_\odot$, $\Delta_{90}(R_{1.4})=1.09~{\rm km}$, and $\Delta_{90}(\Lambda_{1.4})=225.0$, compared with the PMN values $0.300\,M_\odot$, $1.01~{\rm km}$, and $196.6$, respectively. These differences are small and indicate comparable credible-region sizes. Overall, both the DDB and RMF-NL analyses show that SBI reproduces the PMN posterior predictions at the level of model parameters, derived nuclear matter properties, and neutron-star observables. The RMF-NL case gives systematically smaller maximum mass, radii, and tidal deformabilities than DDB, reflecting a softer posterior EoS, but in both parametrizations, the SBI inference remains statistically consistent with the full nested-sampling benchmark.

\begin{table}[t]
    \centering
    \caption{
    Combined comparison of the density-dependent RMF (DDB)and RMF-NL results obtained from SBI and PyMultiNest. 
    Values denote posterior medians with 90\% credible intervals.
    }
    \label{tab:ddrmf_rmfnl_combined}
    \renewcommand{\arraystretch}{1.08}
    \setlength{\tabcolsep}{3.5pt}
    \scriptsize

    \begin{tabular}{lcc|lcc}
        \hline
        \multicolumn{3}{c|}{\textit{DDB}} 
        & \multicolumn{3}{c}{\textit{RMF-NL}} \\
        Quantity & SBI & PMN & Quantity & SBI & PMN \\
        \hline
        $a_\sigma$ 
        & $0.081^{+0.061}_{-0.041}$ 
        & $0.087^{+0.121}_{-0.065}$ 
        & $\Gamma_\sigma$ 
        & $8.20^{+0.560}_{-0.422}$ 
        & $8.15^{+0.678}_{-0.407}$ \\

        $a_\omega$ 
        & $0.040^{+0.050}_{-0.032}$ 
        & $0.047^{+0.073}_{-0.042}$ 
        & $\Gamma_\omega$ 
        & $9.40^{+0.988}_{-0.763}$ 
        & $9.22^{+1.277}_{-0.735}$ \\

        $a_\rho$ 
        & $0.548^{+0.455}_{-0.334}$ 
        & $0.534^{+0.469}_{-0.431}$ 
        & $\Gamma_\rho$ 
        & $11.51^{+2.33}_{-2.45}$ 
        & $11.53^{+2.10}_{-2.24}$ \\

        $\Gamma_{\sigma,0}$ 
        & $9.053^{+0.903}_{-0.819}$ 
        & $9.068^{+1.156}_{-1.431}$ 
        & $\kappa$ 
        & $0.0552^{+0.0247}_{-0.0268}$ 
        & $0.0607^{+0.0207}_{-0.0267}$ \\

        $\Gamma_{\omega,0}$ 
        & $10.800^{+1.348}_{-1.262}$ 
        & $10.835^{+1.733}_{-2.310}$ 
        & $\lambda_0$ 
        & $-0.0047^{+0.0462}_{-0.0372}$ 
        & $-0.0073^{+0.0485}_{-0.0338}$ \\

        $\Gamma_{\rho,0}$ 
        & $8.026^{+0.806}_{-0.933}$ 
        & $7.711^{+0.828}_{-0.890}$ 
        & $\zeta$ 
        & $0.0047^{+0.0135}_{-0.0043}$ 
        & $0.0044^{+0.0127}_{-0.0040}$ \\

        -- & -- & --
        & $\Lambda_\omega$ 
        & $0.0851^{+0.0468}_{-0.0403}$ 
        & $0.0896^{+0.0473}_{-0.0406}$ \\
        \hline
    \end{tabular}

    \vspace{0.6em}

    \begin{tabular}{lcc|cc}
        \hline
        \multirow{2}{*}{Quantity} 
        & \multicolumn{2}{c|}{\textit{DDB}} 
        & \multicolumn{2}{c}{\textit{RMF-NL}} \\
        & SBI & PMN & SBI & PMN \\
        \hline

        $\rho_0$ [fm$^{-3}$] 
        & $0.1522^{+0.0080}_{-0.0071}$ 
        & $0.1528^{+0.0055}_{-0.0055}$ 
        & $0.1519^{+0.0080}_{-0.0083}$ 
        & $0.1524^{+0.0068}_{-0.0064}$ \\

        $E_0$ [MeV] 
        & $-16.18^{+0.53}_{-0.50}$ 
        & $-16.10^{+0.32}_{-0.32}$ 
        & $-16.00^{+0.33}_{-0.33}$ 
        & $-16.10^{+0.30}_{-0.30}$ \\

        $K_0$ [MeV] 
        & $243.9^{+53.6}_{-48.9}$ 
        & $239.8^{+52.7}_{-57.7}$ 
        & $251.1^{+45.6}_{-40.1}$ 
        & $247.9^{+37.1}_{-36.1}$ \\

        $J_{\rm sym}$ [MeV] 
        & $33.45^{+2.91}_{-3.22}$ 
        & $32.30^{+2.79}_{-2.84}$ 
        & $32.02^{+2.80}_{-2.79}$ 
        & $32.09^{+2.53}_{-2.54}$ \\

        $L_{\rm sym}$ [MeV] 
        & $44.3^{+35.1}_{-52.1}$ 
        & $46.1^{+42.8}_{-57.2}$ 
        & $40.9^{+21.9}_{-15.3}$ 
        & $40.5^{+20.5}_{-14.3}$ \\

        $K_{\rm sym}$ [MeV] 
        & $-97.1^{+101.0}_{-46.1}$ 
        & $-83.8^{+99.4}_{-50.6}$ 
        & $-151.5^{+56.4}_{-44.6}$ 
        & $-155.8^{+62.3}_{-43.4}$ \\

        \hline
        $M_{\max}$ [$M_\odot$] 
        & $2.176^{+0.175}_{-0.104}$ 
        & $2.129^{+0.265}_{-0.211}$ 
        & $1.996^{+0.152}_{-0.163}$ 
        & $1.982^{+0.149}_{-0.151}$ \\

        $R(M_{\max})$ [km] 
        & $11.32^{+0.61}_{-0.57}$ 
        & $11.30^{+0.64}_{-0.76}$ 
        & $10.45^{+0.54}_{-0.44}$ 
        & $10.36^{+0.53}_{-0.37}$ \\

        $R_{1.4}$ [km] 
        & $12.82^{+0.70}_{-0.64}$ 
        & $12.77^{+0.85}_{-0.80}$ 
        & $11.82^{+0.54}_{-0.55}$ 
        & $11.74^{+0.49}_{-0.52}$ \\

        $\Lambda_{1.4}$ 
        & $516.0^{+209.6}_{-151.9}$ 
        & $521.1^{+225.8}_{-183.9}$ 
        & $378.0^{+128.8}_{-96.2}$ 
        & $360.5^{+112.0}_{-84.6}$ \\
        \hline
    \end{tabular}
\end{table}

\subsection{Mock-observation test at fixed $R_{1.4}$}

\begin{figure}[t]
    \centering
    \includegraphics[width=0.99\columnwidth]{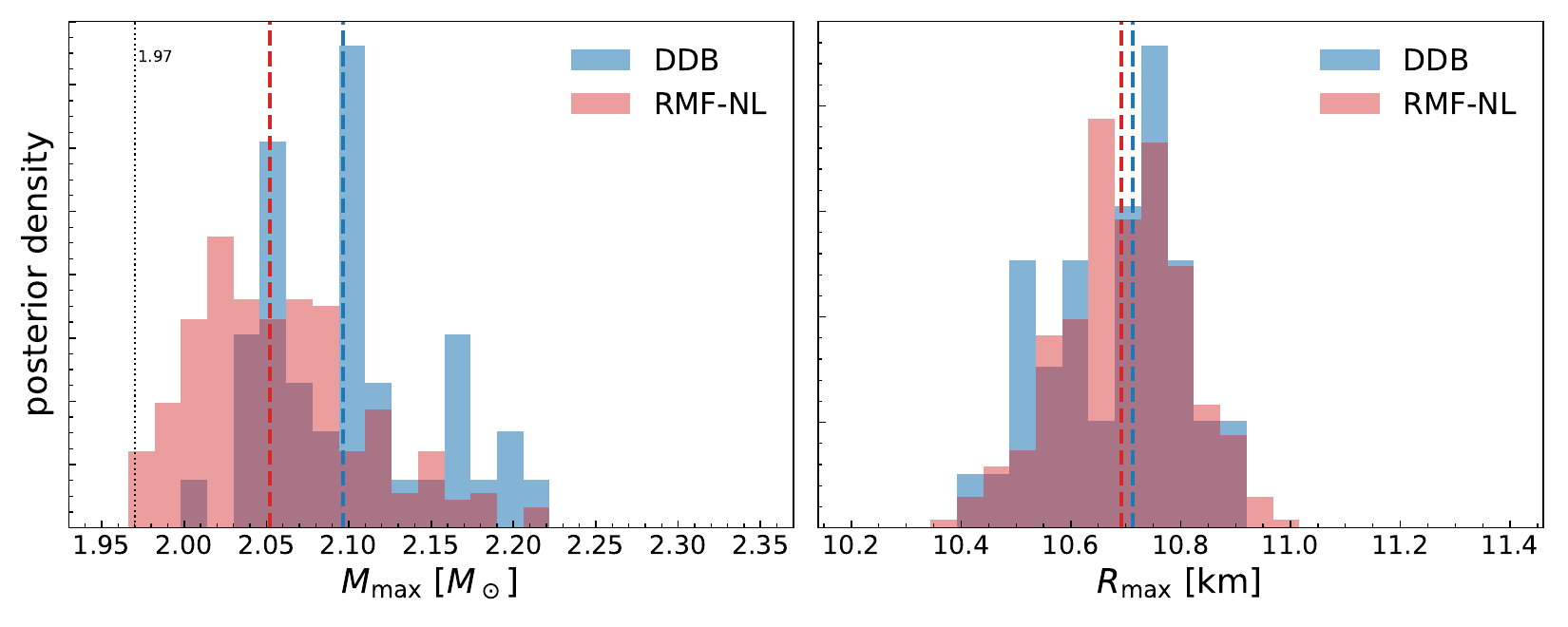}
    \caption{Posterior $M_{\max}$ (left) and $R_{\max}=R(M_{\max})$ (right) for EoS
  samples with $R_{1.4}=12$~km and $M_{\max}>1.97\,M_\odot$, for DDB (blue) and
  RMF-NL (red). Dashed lines mark the medians; the dotted line is the
  $1.97\,M_\odot$ threshold.}
    \label{fig:fixed_radius}
\end{figure}

As an additional, controlled test of the trained estimators, we impose on both families a common mock observation that fixes the canonical radius to $R_{1.4}=12\,{\rm km}$ and imposes $M_{\rm max}>1.97\,M_\odot$, and we examine the resulting maximum-mass configuration. Anchoring the EoS at $1.4\,M_\odot$ removes most of the freedom in the intermediate-density regime and confines both DDB and RMF-NL to a narrow EoS region, so that any residual difference between the two families is driven almost entirely by their high-density behavior. Figure~\ref{fig:fixed_radius} shows the resulting posterior distributions of $M_{\rm max}$ and $R_{\rm max}=R(M_{\rm max})$. The two families give fully consistent predictions: DDB yields $M_{\rm max}=2.10^{+0.09}_{-0.07}\,M_\odot$ and $R_{\rm max}=10.71^{+0.14}_{-0.21}\,{\rm km}$, while RMF-NL gives $M_{\rm max}=2.05^{+0.10}_{-0.06}\,M_\odot$ and $R_{\rm max}=10.69^{+0.18}_{-0.19}\,{\rm km}$. The radius of the maximum-mass star is essentially identical in the two cases, whereas the maximum mass is marginally larger for DDB. This reflects the fact that, once $R_{1.4}$ is fixed, RMF-NL remains slightly softer than DDB at high density and therefore supports a slightly lower maximum mass, consistent with the EoS and mass--radius trends found above.

\subsection{Amortized Neural Posterior Estimation}

\begin{figure}[t]
    \centering
    \includegraphics[width=0.65\columnwidth]{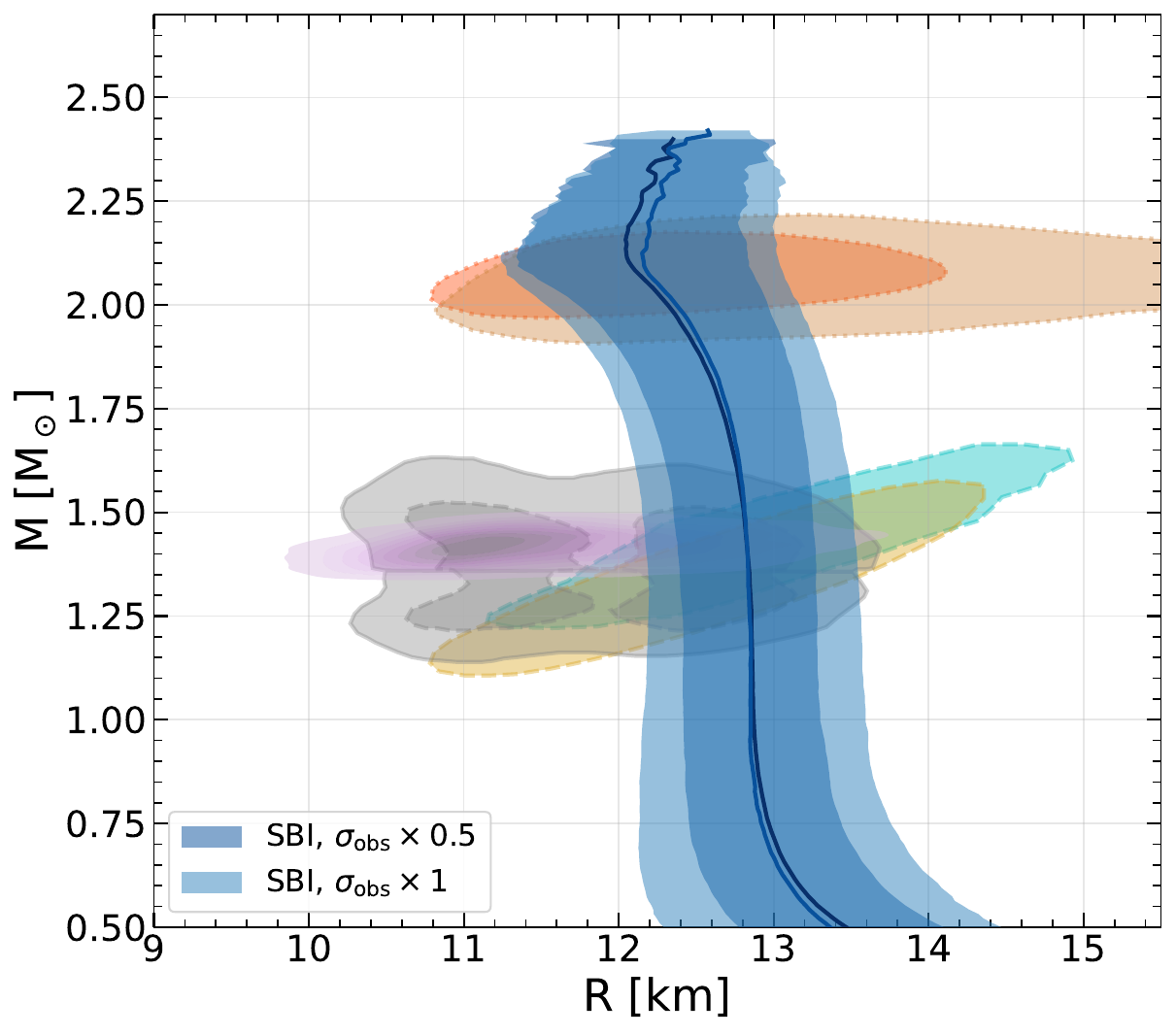}
    \caption{Mass--radius posterior contours illustrating the amortized nature of the SBI framework. After a single training stage, the same neural posterior estimator is evaluated for two input-uncertainty configurations without retraining. The $\sigma_{\rm obs}\times 0.5$ configuration uses half of the base uncertainties, while $\sigma_{\rm obs}\times 1$ corresponds to the base $\sigma$ values listed in Table~\ref{tab:fiducial_input}. The mass--radius contours represent the 90\% credible regions, and the solid curves denote the corresponding median mass--radius relations.}
    \label{fig:amortization_mr}
\end{figure}

Figure~\ref{fig:amortization_mr} illustrates the amortized nature of the neural posterior estimator through the mass--radius posterior. The same trained network is conditioned on the central input vector listed in Table~\ref{tab:fiducial_input}, using either the base uncertainties, $\sigma_{\rm obs}\times1$, or uncertainties reduced by a factor of two, $\sigma_{\rm obs}\times0.5$, without retraining. Reducing the input uncertainties contracts the mass--radius posterior while leaving its median relation nearly unchanged. For $\sigma_{\rm obs}\times0.5$, we obtain
$M_{\rm max}=2.179^{+0.159}_{-0.089}\,M_\odot$,
$R(M_{\rm max})=11.29^{+0.48}_{-0.41}\,\mathrm{km}$, and
$R_{1.4}=12.83^{+0.44}_{-0.45}\,\mathrm{km}$, compared with
$M_{\rm max}=2.179^{+0.179}_{-0.099}\,M_\odot$,
$R(M_{\rm max})=11.33^{+0.60}_{-0.57}\,\mathrm{km}$, and
$R_{1.4}=12.83^{+0.71}_{-0.65}\,\mathrm{km}$ for the baseline
$\sigma_{\rm obs}\times1$ case. Thus, after the one-time training cost, the posterior can be updated for different uncertainty assumptions without repeating a likelihood-based scan. A posterior containing $3\times10^{4}$ samples is generated in approximately $2.5$~s on a CPU, whereas nested sampling must be rerun for each modified input or uncertainty model.

\section{Summary and conclusions}
\label{Summary}
We developed an amortized simulation-based inference framework for constraining the microscopic couplings of two RMF EoS families, DDB and RMF-NL. Using broad prior-predictive training banks, the neural posterior estimator successfully learned the mapping between the RMF couplings, nuclear-matter inputs, the EoS, and neutron-star observables. The broad priors were thereby reduced to well-constrained posterior regions. The inferred posteriors agree well with the one obtained with a conventional inference technique, i.e, nested sampling with the PyMultiNest sampler. For DDB, the two-sample Kolmogorov--Smirnov test~\cite{Massey01031951} gives $D=0.065$ for $R_{1.4}$, $D=0.038$ for $\Lambda_{1.4}$, and $D=0.075$ for $R(M_{\max})$, indicating close agreement between the SBI and PMN posterior predictions. These values correspond to the specified fixed-mass and maximum-mass configurations, rather than to the full mass range. The larger value, $D=0.282$ for $M_{\max}$, reflects the slightly stiffer high-density EoSs favored by SBI. For RMF-NL, the corresponding statistics are $D=0.084$ for $M_{\max}$, $D=0.135$ for $R_{1.4}$, $D=0.138$ for $\Lambda_{1.4}$, and $D=0.149$ for $R(M_{\max})$, showing only modest differences between the SBI and PMN posteriors.
The main advantage is amortization: after the one-time simulation and training cost, posterior samples are generated within seconds, and modified uncertainty assumptions can be evaluated without retraining or repeating a likelihood-based scan. Constraints from NICER mass-radii likelihoods and GWs were not included directly in the present conditioning vector. This choice was deliberate: the primary objective was to perform a controlled, one-to-one validation of microscopic RMF-coupling inference against PyMultiNest using an identical, compact set of nuclear and $M_{\rm max}$. Future work will extend the framework beyond the present low-dimensional conditioning vector by incorporating multimessenger observations more directly, including NICER mass-radii measurements information and gravitational-wave tidal likelihoods. Such extensions would retain the amortized character of the method while enabling \red{live} rapid, joint microscopic-EoS inference as new neutron-star observations and revised measurement uncertainties become available.

\section*{Acknowledgements}
P.~Thakur is supported by the National Research Foundation of Korea (NRF) grant funded by the Korea government (MSIT) (No.~RS-2024-00457037). This work was supported (in part) by the Yonsei University Research Fund(Yonsei University Frontier Fellowship for Postdoctoral Researchers) of 2025. T.M would like to thank the national funds from FCT (Fundação para a Ciência e a Tecnologia, I.P, Portugal) under project UID/04564/2025, identified by DOI 10.54499/UIDB/04564/2025, and project  2024.16290.PEX identified by DOI  identifier 10.54499/2024.16290.PEX.

\section*{Data availability}

The numerical data required to reproduce all figures presented in this work will be made publicly available in a repository upon acceptance of the manuscript. The simulation-based inference analysis was performed using the publicly available \texttt{sbi} package with neural posterior estimation, and the training setup, input observables, prior ranges, network configuration, and computational details are specified in the manuscript.

\appendix

\section{Simulation-Based Inference and Training Setup}
\label{app:sbi_training}

\subsection{Data generation}
\label{app:data_generation}

Before training the neural posterior estimator (NPE), we construct
prior-predictive training banks for both RMF models considered in this work,
namely DDB and RMF-NL. The data-generation strategy is the same in both
cases: model parameters are sampled from their corresponding prior ranges,
evolved through the respective forward pipeline, and the resulting
nuclear-matter and neutron-star observables are stored as the conditioning
vector \(\mathbf{x}\). Since the construction is identical in structure for
the two models, we describe it explicitly for the representative DDB case.

For the DDB model, the parameter vector contains the six density-dependent
coupling parameters,
\begin{equation}
\boldsymbol{\theta}
=
\left(
a_\sigma,\,
a_\omega,\,
a_\rho,\,
\Gamma_{\sigma,0},\,
\Gamma_{\omega,0},\,
\Gamma_{\rho,0}
\right),
\end{equation}
which are sampled uniformly from broad flat priors. For each sampled
parameter vector \(\boldsymbol{\theta}\), the DDB model is evaluated through
the forward pipeline to obtain the EoS, the relevant nuclear-matter
observables, the PNM pressures, and the maximum mass from the TOV equations.
The resulting training bank contains \(2{,}129{,}225\) accepted
\((\boldsymbol{\theta},\mathbf{x})\) pairs, with
\begin{equation}
\begin{aligned}
\mathbf{x}
=
\big(&
\rho_0,\,
E_0,\,
K_0,\,
J_{\rm sym}, \\
&
P_{\rm PNM}(0.08),\,
P_{\rm PNM}(0.12),\,
P_{\rm PNM}(0.16),\,
M_{\rm max}
\big).
\end{aligned}
\end{equation}

The sampling ranges are deliberately chosen to cover a broad
prior-predictive domain rather than only the region close to the final
empirical constraints. In particular, the bank spans approximately
\(E_0\in[-100,100]\) MeV, \(K_0\lesssim 800\) MeV, and PNM pressures
extending beyond the fiducial chiral-EFT values used later for conditioning.
During bank construction, we impose only the loose acceptance condition
\(M_{\rm max}\geq 1.4\,M_\odot\), which removes extremely soft models but is
not intended as a physical lower bound on neutron-star masses. In the
accepted DDB bank, \(M_{\rm max}\) spans
\(1.40\)--\(3.56\,M_\odot\), with a median value of approximately
\(2.28\,M_\odot\). The corresponding prior and prior-predictive ranges for
both DDB and RMF-NL are summarized in
Table 1.

\subsection{Neural Posterior Estimation}
\label{app:npe_training}

We train NPE~\cite{papamakarios2018fastepsilonfreeinferencesimulation}
using the \texttt{sbi} package~\cite{BoeltsDeistler_sbi_2025}. Neural
posterior estimation directly learns an approximation to the posterior
distribution of the model parameters conditioned on the simulated
observables. In the present case, the network learns
\begin{equation}
p(\boldsymbol{\theta}\mid\mathbf{x})
\simeq
q_{\phi}(\boldsymbol{\theta}\mid\mathbf{x}),
\end{equation}
where \(\boldsymbol{\theta}\) denotes the DDB model parameters,
\(\mathbf{x}\) denotes the conditioning observables, and \(q_{\phi}\) is the
neural density estimator with trainable weights \(\phi\). The estimator is
trained by maximum likelihood on the simulated pairs
\((\boldsymbol{\theta}_i,\mathbf{x}_i)\), minimizing the negative
log-probability that the flow assigns to each parameter set given its
observables,
\begin{equation}
{\cal L}(\phi)
=
-\sum_i
\log q_{\phi}
\left(
\boldsymbol{\theta}_i\mid\mathbf{x}_i
\right).
\end{equation}

As illustrated in Fig.1 
, which depicts the SBI training
pipeline, after constructing a large prior-predictive training dataset, we
train an NPE,
\(q_{\phi}(\boldsymbol{\theta}\mid\mathbf{x})\), to infer the model
parameters \(\boldsymbol{\theta}\) from the simulated observable vector
\(\mathbf{x}\). The parameter vector consists of six couplings for the DDB
model and seven couplings for the RMF-NL model. The conditional posterior
density is modeled using a normalizing-flow architecture, specifically a
neural spline flow (NSF) with \(20\)--\(25\) coupling transforms and
\(256\)--\(320\) hidden features. The trained density-estimator state
dictionary contains 554 tensors. The main training hyperparameters are a
batch size of 2048, a learning rate of \(3\times10^{-4}\), a maximum of
500 epochs, validation evaluation every 10 epochs, and an early-stopping
patience of 120 epochs. We use the Adam optimizer with the default
implementation in \texttt{sbi}.

Training is performed in successive 10-epoch chunks using
\texttt{resume\_training=True}. After each chunk, we carry out an additional
custom validation step using up to 512 held-out validation points, drawing
2048 posterior samples for each point. Model selection is based on a
composite validation score,
\begin{equation}
{\cal S}
=
{\rm NLPD}
+
5\,\left|C_{68}-0.68\right|
+
5\,\left|C_{90}-0.90\right|
+
0.2\,{\rm MAE},
\end{equation}
where \(C_{68}\) and \(C_{90}\) denote the empirical 68\% and 90\% coverage
fractions, respectively, NLPD is the negative log-posterior density, and MAE
is the mean absolute error of the posterior location. These settings are
treated as training hyperparameters and require careful tuning, since the
optimal configuration depends on the size of the training bank, the
validation performance, and the posterior calibration score.

All NPE training was performed on an NVIDIA GeForce RTX 3060 Ti GPU with
8 GB GDDR6 memory and Ampere compute capability 8.6. The software stack used
PyTorch 2.5.1 built against CUDA 12.4 with cuDNN 9.2.1, using driver
version 580.159.03. The one-time training cost for a full 500-epoch run was
approximately \(2.2\) h, corresponding to about \(15.8\) s per epoch. Once
trained, posterior sampling is fast: drawing \(4\times10^{5}\) NPE
posterior samples required approximately \(5.6\) s.

\subsection{Noise conditioning}
\label{app:noise_conditioning}

A key element of the training phase is noise conditioning. For each
simulated observable vector \(\mathbf{x}\), we generate several
noise-augmented replicas using an uncertainty scale
\begin{equation}
\boldsymbol{\sigma}
=
m\,\boldsymbol{\sigma}_{\rm base},
\qquad
m\sim\mathcal{U}(0.5,2.0),
\end{equation}
where \(\boldsymbol{\sigma}_{\rm base}\) denotes the fiducial observational
uncertainty vector described in Table 2 
The corresponding noise scale \(\boldsymbol{\sigma}\) is supplied to the neural
posterior estimator as an additional conditioning input. This exposes the
normalizing flow to a continuous range of measurement uncertainties rather
than a single fixed noise level.

As a result, the estimator learns posteriors that remain robust across
different observational precisions and avoids the overly narrow,
overconfident posteriors that amortized flows can otherwise produce. In this
sense, the inference is amortized not only over the observable vector
\(\mathbf{x}\), but also over the uncertainty scale
\(\boldsymbol{\sigma}\). A single trained network can therefore return
calibrated posterior samples over the full noise range without retraining.

The simulations are divided using a \(90/10\) train--validation split, and
statistical calibration is assessed only on the held-out validation set
using the TARP diagnostic.

\subsection{Posterior calibration and processing}
\label{app:posterior_calibration}

At inference, the trained flow is evaluated at the fiducial observation
vector \(\mathbf{x}_{\rm obs}\), as given in
Table 2 
, yielding the raw posterior approximation
\begin{equation}
p(\boldsymbol{\theta}\mid\mathbf{x}_{\rm obs})
\simeq
q_{\phi}(\boldsymbol{\theta}\mid\mathbf{x}_{\rm obs}).
\end{equation}

The raw posterior is then calibrated in both spread and location. Where
indicated by the TARP diagnostic, a per-parameter temperature factor
\(T_{\star}\) is applied to restore correct posterior coverage, motivated
by the known tendency of neural simulation-based posterior estimators to
produce overconfident
posteriors~\cite{hermans2022trustcrisissimulationbasedinference}. The need
for this correction is assessed using the TARP coverage
diagnostic~\cite{2023PMLR..20219256L}. This correction is used for the
RMF-NL model, while the raw DDB posterior already satisfies the TARP
calibration test.

In addition, a small sub-\(\sigma\) location correction, diagnosed from the
held-out validation set, is applied to remove residual posterior-location
bias while preserving the posterior widths and covariance structure. The
calibrated samples are subsequently passed through an accept--reject step
imposing the astrophysical constraint
\begin{equation}
M_{\max}>2\,M_{\odot}.
\end{equation}

The resulting final posterior,
\begin{equation}
p(\boldsymbol{\theta}\mid\mathbf{x}_{\rm obs}),
\end{equation}
is propagated through the same forward model to obtain the equation of state
\(P(\rho)\), the mass--radius and mass--tidal-deformability bands, and the
derived tables of neutron-star observables and nuclear-matter properties.
These final SBI-based results are then compared with an independent Bayesian
reference obtained using nested sampling.

\bibliographystyle{elsarticle-num}
\bibliography{example}

@ARTICLE{2023PMLR..20219256L,
       author = {{Lemos}, Pablo and {Coogan}, Adam and {Hezaveh}, Yashar and {Perreault-Levasseur}, Laurence},
        title = "{Sampling-Based Accuracy Testing of Posterior Estimators for General Inference}",
      journal = {40th International Conference on Machine Learning},
         year = 2023,
        month = jan,
       volume = {202},
        pages = {19256-19273},
          doi = {10.48550/arXiv.2302.03026},
archivePrefix = {arXiv},
       eprint = {2302.03026},
 primaryClass = {stat.ML},
       adsurl = {https://ui.adsabs.harvard.edu/abs/2023PMLR..20219256L}
}

@article{BoeltsDeistler_sbi_2025,
  doi = {10.21105/joss.07754},
  url = {https://doi.org/10.21105/joss.07754},
  year = {2025},
  publisher = {The Open Journal},
  volume = {10},
  number = {108},
  pages = {7754},
  author = {Jan Boelts and Michael Deistler and Manuel Gloeckler and Álvaro Tejero-Cantero and Jan-Matthis Lueckmann and Guy Moss and Peter Steinbach and Thomas Moreau and Fabio Muratore and Julia Linhart and Conor Durkan and Julius Vetter and Benjamin Kurt Miller and Maternus Herold and Abolfazl Ziaeemehr and Matthijs Pals and Theo Gruner and Sebastian Bischoff and Nastya Krouglova and Richard Gao and Janne K. Lappalainen and Bálint Mucsányi and Felix Pei and Auguste Schulz and Zinovia Stefanidi and Pedro Rodrigues and Cornelius Schröder and Faried Abu Zaid and Jonas Beck and Jaivardhan Kapoor and David S. Greenberg and Pedro J. Gonçalves and Jakob H. Macke},
  title = {sbi reloaded: a toolkit for simulation-based inference workflows},
  journal = {Journal of Open Source Software}
}

@article{Miller:2021qha,
    author = "Miller, M. C. and others",
    title = "{The Radius of PSR J0740+6620 from NICER and XMM-Newton Data}",
    eprint = "2105.06979",
    archivePrefix = "arXiv",
    primaryClass = "astro-ph.HE",
    doi = "10.3847/2041-8213/ac089b",
    journal = "Astrophys. J. Lett.",
    volume = "918",
    number = "2",
    pages = "L28",
    year = "2021"
}

@article{Choudhury:2024xbk,
    author = "Choudhury, Devarshi and others",
    title = "{A NICER View of the Nearest and Brightest Millisecond Pulsar: PSR J0437\textendash{}4715}",
    doi = "10.3847/2041-8213/ad5a6f",
    journal = "Astrophys. J. Lett.",
    volume = "971",
    number = "1",
    pages = "L20",
    year = "2024"
}

@article{malik2022,
    author = "Malik, Tuhin and Ferreira, M\'arcio and Agrawal, B. K. and Provid\^encia, Constan\c{c}a",
    title = "{Relativistic Description of Dense Matter Equation of State and Compatibility with Neutron Star Observables: A Bayesian Approach}",
    doi = "10.3847/1538-4357/ac5d3c",
    journal = "Astrophys. J.",
    volume = "930",
    number = "1",
    pages = "17",
    year = "2022"
}

@article{Riley:2021pdl,
    author = "Riley, Thomas E. and others",
    title = "{A NICER View of the Massive Pulsar PSR J0740+6620 Informed by Radio Timing and XMM-Newton Spectroscopy}",
    eprint = "2105.06980",
    archivePrefix = "arXiv",
    primaryClass = "astro-ph.HE",
    doi = "10.3847/2041-8213/ac0a81",
    journal = "Astrophys. J. Lett.",
    volume = "918",
    number = "2",
    pages = "L27",
    year = "2021"
}

@article{Essick:2021kjb,
    author = "Essick, Reed and Tews, Ingo and Landry, Philippe and Schwenk, Achim",
    title = "{Astrophysical Constraints on the Symmetry Energy and the Neutron Skin of Pb208 with Minimal Modeling Assumptions}",
    eprint = "2102.10074",
    archivePrefix = "arXiv",
    primaryClass = "nucl-th",
    reportNumber = "LA-UR-21-20527",
    doi = "10.1103/PhysRevLett.127.192701",
    journal = "Phys. Rev. Lett.",
    volume = "127",
    number = "19",
    pages = "192701",
    year = "2021"
}

@article{Annala:2019puf,
    author = "Annala, Eemeli and Gorda, Tyler and Kurkela, Aleksi and Nättilä, Joonas and Vuorinen, Aleksi",
    title = "{Evidence for quark-matter cores in massive neutron stars}",
    eprint = "1903.09121",
    archivePrefix = "arXiv",
    primaryClass = "astro-ph.HE",
    reportNumber = "CERN-TH-2019-031, HIP-2019-7/TH",
    doi = "10.1038/s41567-020-0914-9",
    journal = "Nature Phys.",
    year = "2020"
}

@article{Miller:2019cac,
    author = "Miller, M.C. and others",
    title = "{PSR J0030+0451 Mass and Radius from $NICER$ Data and Implications for the Properties of Neutron Star Matter}",
    eprint = "1912.05705",
    archivePrefix = "arXiv",
    primaryClass = "astro-ph.HE",
    doi = "10.3847/2041-8213/ab50c5",
    journal = "Astrophys. J. Lett.",
    volume = "887",
    number = "1",
    pages = "L24",
    year = "2019"
}

@article{Riley:2019yda,
    author = "Riley, Thomas E. and others",
    title = "{A $NICER$ View of PSR J0030+0451: Millisecond Pulsar Parameter Estimation}",
    eprint = "1912.05702",
    archivePrefix = "arXiv",
    primaryClass = "astro-ph.HE",
    doi = "10.3847/2041-8213/ab481c",
    journal = "Astrophys. J. Lett.",
    volume = "887",
    number = "1",
    pages = "L21",
    year = "2019"
}

@article{Hebeler:2013nza,
    author = "Hebeler, K. and Lattimer, J.M. and Pethick, C.J. and Schwenk, A.",
    title = "{Equation of state and neutron star properties constrained by nuclear physics and observation}",
    eprint = "1303.4662",
    archivePrefix = "arXiv",
    primaryClass = "astro-ph.SR",
    doi = "10.1088/0004-637X/773/1/11",
    journal = "Astrophys. J.",
    volume = "773",
    pages = "11",
    year = "2013"
}

@article{Burgio:2021vgk,
    author = "Burgio, G. F. and Schulze, H. -J. and Vidana, I. and Wei, J. -B.",
    title = "{Neutron stars and the nuclear equation of state}",
    eprint = "2105.03747",
    archivePrefix = "arXiv",
    primaryClass = "nucl-th",
    doi = "10.1016/j.ppnp.2021.103879",
    journal = "Prog. Part. Nucl. Phys.",
    volume = "120",
    pages = "103879",
    year = "2021"
}

@article{Tews:2012fj,
    author = {Tews, I. and Kr\"uger, T. and Hebeler, K. and Schwenk, A.},
    title = "{Neutron matter at next-to-next-to-next-to-leading order in chiral effective field theory}",
    eprint = "1206.0025",
    archivePrefix = "arXiv",
    primaryClass = "nucl-th",
    doi = "10.1103/PhysRevLett.110.032504",
    journal = "Phys. Rev. Lett.",
    volume = "110",
    number = "3",
    pages = "032504",
    year = "2013"
}

@article{Raaijmakers:2021uju,
    author = "Raaijmakers, G. and Greif, S. K. and Hebeler, K. and Hinderer, T. and Nissanke, S. and Schwenk, A. and Riley, T. E. and Watts, A. L. and Lattimer, J. M. and Ho, W. C. G.",
    title = "{Constraints on the Dense Matter Equation of State and Neutron Star Properties from NICER\textquoteright{}s Mass\textendash{}Radius Estimate of PSR J0740+6620 and Multimessenger Observations}",
    eprint = "2105.06981",
    archivePrefix = "arXiv",
    primaryClass = "astro-ph.HE",
    doi = "10.3847/2041-8213/ac089a",
    journal = "Astrophys. J. Lett.",
    volume = "918",
    number = "2",
    pages = "L29",
    year = "2021"
}

@Article{Landry:2020vaw,
  author        = {Landry, Philippe and Essick, Reed and Chatziioannou, Katerina},
  title         = {{Nonparametric constraints on neutron star matter with existing and upcoming gravitational wave and pulsar observations}},
  doi           = {10.1103/PhysRevD.101.123007},
  eprint        = {2003.04880},
  number        = {12},
  pages         = {123007},
  volume        = {101},
  archiveprefix = {arXiv},
  journal       = {Phys. Rev. D},
  primaryclass  = {astro-ph.HE},
  year          = {2020},
}

@Article{Dutra:2014qga,
  author        = {Dutra, M. and Louren\c{c}o, O. and Avancini, S. S. and Carlson, B. V. and Delfino, A. and Menezes, D. P. and Provid\^encia, C. and Typel, S. and Stone, J. R.},
  title         = {{Relativistic Mean-Field Hadronic Models under Nuclear Matter Constraints}},
  doi           = {10.1103/PhysRevC.90.055203},
  eprint        = {1405.3633},
  number        = {5},
  pages         = {055203},
  volume        = {90},
  archiveprefix = {arXiv},
  journal       = {Phys. Rev. C},
  primaryclass  = {nucl-th},
  year          = {2014},
}

@article{Steiner:2010fz,
    author = "Steiner, Andrew W. and Lattimer, James M. and Brown, Edward F.",
    title = "{The Equation of State from Observed Masses and Radii of Neutron Stars}",
    eprint = "1005.0811",
    archivePrefix = "arXiv",
    primaryClass = "astro-ph.HE",
    doi = "10.1088/0004-637X/722/1/33",
    journal = "Astrophys. J.",
    volume = "722",
    pages = "33--54",
    year = "2010"
}

@article{Huth:2021bsp,
    author = "Huth, S. and others",
    title = "{Constraining Neutron-Star Matter with Microscopic and Macroscopic Collisions}",
    eprint = "2107.06229",
    archivePrefix = "arXiv",
    primaryClass = "nucl-th",
    reportNumber = "LA-UR-21-22072",
    doi = "10.1038/s41586-022-04750-w",
    journal = "Nature",
    volume = "606",
    pages = "276--280",
    year = "2022"
}

@article{Komoltsev:2021jzg,
    author = "Komoltsev, Oleg and Kurkela, Aleksi",
    title = "{How Perturbative QCD Constrains the Equation of State at Neutron-Star Densities}",
    eprint = "2111.05350",
    archivePrefix = "arXiv",
    primaryClass = "nucl-th",
    doi = "10.1103/PhysRevLett.128.202701",
    journal = "Phys. Rev. Lett.",
    volume = "128",
    number = "20",
    pages = "202701",
    year = "2022"
}

@article{Fonseca:2016tux,
    author = "Fonseca, Emmanuel and others",
    title = "{The NANOGrav Nine-year Data Set: Mass and Geometric Measurements of Binary Millisecond Pulsars}",
    eprint = "1603.00545",
    archivePrefix = "arXiv",
    primaryClass = "astro-ph.HE",
    doi = "10.3847/0004-637X/832/2/167",
    journal = "Astrophys. J.",
    volume = "832",
    number = "2",
    pages = "167",
    year = "2016"
}

@article{Reitze:2019iox,
    author = "Reitze, David and others",
    title = "{Cosmic Explorer: The U.S. Contribution to Gravitational-Wave Astronomy beyond LIGO}",
    eprint = "1907.04833",
    archivePrefix = "arXiv",
    primaryClass = "astro-ph.IM",
    reportNumber = "LIGO-P1900316",
    journal = "Bull. Am. Astron. Soc.",
    volume = "51",
    number = "7",
    pages = "035",
    year = "2019"
}

@article{Raaijmakers:2019dks,
    author = "Raaijmakers, G. and others",
    title = "{Constraining the dense matter equation of state with joint analysis of NICER and LIGO/Virgo measurements}",
    eprint = "1912.11031",
    archivePrefix = "arXiv",
    primaryClass = "astro-ph.HE",
    doi = "10.3847/2041-8213/ab822f",
    journal = "Astrophys. J. Lett.",
    volume = "893",
    number = "1",
    pages = "L21",
    year = "2020"
}

@article{LIGOScientific:2017vwq,
    author = "Abbott, B. P. and others",
    collaboration = "LIGO Scientific, Virgo",
    title = "{GW170817: Observation of Gravitational Waves from a Binary Neutron Star Inspiral}",
    eprint = "1710.05832",
    archivePrefix = "arXiv",
    primaryClass = "gr-qc",
    reportNumber = "LIGO-P170817",
    doi = "10.1103/PhysRevLett.119.161101",
    journal = "Phys. Rev. Lett.",
    volume = "119",
    number = "16",
    pages = "161101",
    year = "2017"
}

@article{Cartaxo:2025jpi,
    author = "Cartaxo, Jo{\~a}o and Huang, Chun and Malik, Tuhin and Sourav, Shashwat and Yuan, Wen-Li and Zhou, Tianzhe and Liu, Xuezhi and Provid{\^e}ncia, Constan{\c{c}}a",
    title = "{Covariant Energy Density Functionals for Modeling the Equation of State of Neutron Star Matter: Cross-comparison Analysis Using CompactObject}",
    eprint = "2506.03112",
    archivePrefix = "arXiv",
    primaryClass = "nucl-th",
    reportNumber = "ET-0331A-25",
    doi = "10.3847/1538-4365/ae2310",
    journal = "Astrophys. J. Suppl.",
    volume = "282",
    number = "2",
    pages = "33",
    year = "2026"
}

@misc{papamakarios2018fastepsilonfreeinferencesimulation,
      title={Fast $\epsilon$-free Inference of Simulation Models with Bayesian Conditional Density Estimation}, 
      author={George Papamakarios and Iain Murray},
      year={2018},
      eprint={1605.06376},
      archivePrefix={arXiv},
      primaryClass={stat.ML},
      url={https://arxiv.org/abs/1605.06376}, 
}

@misc{hermans2022trustcrisissimulationbasedinference,
      title={A Trust Crisis In Simulation-Based Inference? Your Posterior Approximations Can Be Unfaithful}, 
      author={Joeri Hermans and Arnaud Delaunoy and François Rozet and Antoine Wehenkel and Volodimir Begy and Gilles Louppe},
      year={2022},
      eprint={2110.06581},
      archivePrefix={arXiv},
      primaryClass={stat.ML},
      url={https://arxiv.org/abs/2110.06581}, 
}

@article{Shlomo:2006ole,
    author = "Shlomo, S. and Kolomietz, V. M. and Col{\`o}, G.",
    title = "{Deducing the nuclear-matter incompressibility coefficient from data on isoscalar compression modes}",
    doi = "10.1140/epja/i2006-10100-3",
    journal = "Eur. Phys. J. A",
    volume = "30",
    number = "1",
    pages = "23--30",
    year = "2006"
}

@article{Typel:1999yq,
    author = "Typel, S. and Wolter, H. H.",
    title = "{Relativistic mean field calculations with density dependent meson nucleon coupling}",
    doi = "10.1016/S0375-9474(99)00310-3",
    journal = "Nucl. Phys. A",
    volume = "656",
    pages = "331--364",
    year = "1999"
}

@article{ET:2025xjr,
    author = "Abac, Adrian and others",
    collaboration = "ET",
    title = "{The Science of the Einstein Telescope}",
    eprint = "2503.12263",
    archivePrefix = "arXiv",
    primaryClass = "gr-qc",
    reportNumber = "ET-0036C-25",
    doi = "10.1088/1475-7516/2026/03/081",
    journal = "JCAP",
    volume = "03",
    pages = "081",
    year = "2026"
}

@article{Chua:2019wwt,
author = {Chua, Alvin J. K. and Vallisneri, Michele},
title = "{Learning Bayesian Posteriors with Neural Networks for Gravitational-Wave Inference}",
eprint = "1909.05966",
archivePrefix = "arXiv",
primaryClass = "gr-qc",
doi = "10.1103/PhysRevLett.124.041102",
journal = "Phys. Rev. Lett.",
volume = "124",
number = "4",
pages = "041102",
year = "2020"
}

@article{Dax:2021tsq,
author = {Dax, Maximilian and Green, Stephen R. and Gair, Jonathan and Macke, Jakob H. and Buonanno, Alessandra and Sch"olkopf, Bernhard},
title = "{Real-Time Gravitational Wave Science with Neural Posterior Estimation}",
eprint = "2106.12594",
archivePrefix = "arXiv",
primaryClass = "gr-qc",
reportNumber = "LIGO-P2100223",
doi = "10.1103/PhysRevLett.127.241103",
journal = "Phys. Rev. Lett.",
volume = "127",
number = "24",
pages = "241103",
year = "2021"
}

@article{Dax:2022pxd,
author = {Dax, Maximilian and Green, Stephen R. and Gair, Jonathan and P"urrer, Michael and Wildberger, Jonas and Macke, Jakob H. and Buonanno, Alessandra and Sch"olkopf, Bernhard},
title = "{Neural Importance Sampling for Rapid and Reliable Gravitational-Wave Inference}",
eprint = "2210.05686",
archivePrefix = "arXiv",
primaryClass = "gr-qc",
doi = "10.1103/PhysRevLett.130.171403",
journal = "Phys. Rev. Lett.",
volume = "130",
number = "17",
pages = "171403",
year = "2023"
}

@article{Dax:2024mcn,
author = {Dax, Maximilian and Green, Stephen R. and Gair, Jonathan and Gupte, Neeraj and P"urrer, Michael and Raymond, Vivien and Wildberger, Jonas and Macke, Jakob H. and Buonanno, Alessandra and Sch"olkopf, Bernhard},
title = "{Real-time inference for binary neutron star mergers using machine learning}",
eprint = "2407.09602",
archivePrefix = "arXiv",
primaryClass = "gr-qc",
doi = "10.1038/s41586-025-08593-z",
journal = "Nature",
volume = "639",
number = "8053",
pages = "49--53",
year = "2025"
}

@article{Brandes:2024abc,
author = {Brandes, Len and Modi, Chirag and Ghosh, Aishik and Farrell, Delaney and Lindblom, Lee and Heinrich, Lukas and Steiner, Andrew W. and Weber, Fridolin and Whiteson, Daniel},
title = "{Neural Simulation-Based Inference of the Neutron Star Equation of State directly from Telescope Spectra}",
eprint = "2403.00287",
archivePrefix = "arXiv",
primaryClass = "astro-ph.HE",
doi = "10.1088/1475-7516/2024/09/009",
journal = "JCAP",
volume = "09",
pages = "009",
year = "2024"
}

@article{Carvalho:2025abc,
    author = {Carvalho, Valeria and Ferreira, Marcio and Bejger, Michal and Providencia, Constanca},
    title = {Neural Posterior Estimation of Neutron Star Equations of State},
    eprint = {2507.23506},
    archivePrefix = {arXiv},
    primaryClass = {nucl-th},
    doi = {10.1103/PhysRevD.112.083044},
    journal = {Phys. Rev. D},
    volume = {112},
    number = {8},
    pages = {083044},
    year = {2025}
}

@article{Lattimer:2006xb,
    author = "Lattimer, James M. and Prakash, Maddapa",
    title = "{Neutron Star Observations: Prognosis for Equation of State Constraints}",
    eprint = "astro-ph/0612440",
    archivePrefix = "arXiv",
    doi = "10.1016/j.physrep.2007.02.003",
    journal = "Phys. Rept.",
    volume = "442",
    pages = "109--165",
    year = "2007"
}

@article{Salmi:2024bss,
    author = "Salmi, Tuomo and others",
    title = "{A NICER View of PSR J1231{\ensuremath{-}}1411: A Complex Case}",
    eprint = "2409.14923",
    archivePrefix = "arXiv",
    primaryClass = "astro-ph.HE",
    doi = "10.3847/1538-4357/ad81d2",
    journal = "Astrophys. J.",
    volume = "976",
    number = "1",
    pages = "58",
    year = "2024"
}

@article{Kurkela:2009gj,
    author = "Kurkela, Aleksi and Romatschke, Paul and Vuorinen, Aleksi",
    title = "{Cold Quark Matter}",
    eprint = "0912.1856",
    archivePrefix = "arXiv",
    primaryClass = "hep-ph",
    reportNumber = "BI-TP-2009-30, CERN-PH-TH-2009-229, INT-PUB-09-060, TUW-09-19",
    doi = "10.1103/PhysRevD.81.105021",
    journal = "Phys. Rev. D",
    volume = "81",
    pages = "105021",
    year = "2010"
}

@article{
doi:10.1073/pnas.1912789117,
author = {Kyle Cranmer  and Johann Brehmer  and Gilles Louppe },
title = {The frontier of simulation-based inference},
journal = {Proceedings of the National Academy of Sciences},
volume = {117},
number = {48},
pages = {30055-30062},
year = {2020},
doi = {10.1073/pnas.1912789117},
URL = {https://www.pnas.org/doi/abs/10.1073/pnas.1912789117},
eprint = {https://www.pnas.org/doi/pdf/10.1073/pnas.1912789117}}

@article{Massey01031951, author = {Frank J. Massey Jr.}, title = {The Kolmogorov-Smirnov Test for Goodness of Fit}, journal = {Journal of the American Statistical Association}, volume = {46}, number = {253}, pages = {68--78}, year = {1951}, publisher = {Taylor \& Francis}, doi = {10.1080/01621459.1951.10500769}, URL = {https://www.tandfonline.com/doi/abs/10.1080/01621459.1951.10500769}, eprint = {https://www.tandfonline.com/doi/pdf/10.1080/01621459.1951.10500769} }






\end{document}